\documentclass[article]{aa} 
\usepackage{graphicx}
\usepackage{txfonts}
\usepackage[colorlinks=true,citecolor=blue]{hyperref}
\usepackage{graphicx}
\usepackage{amsmath,amssymb}
\usepackage{times}
\usepackage{natbib}
\usepackage{color}
\usepackage{multirow}
\usepackage{todonotes}
\usepackage{appendix}
\usepackage{array}
\usepackage{verbatim}
\usepackage{enumitem}
\usepackage{pifont}

\definecolor{DarkBlue}{rgb}{0.1,0.2,0.7} 
\newcommand{\msol}{\rm M_{\odot}}
\newcommand{\pkin}{\dot{E}_{\mathrm{kin}}}

\newcommand{\fkin}{f_{\mathrm{kin}}}
\newcommand{\Pjet}{P_{\rm jet}}
\newcommand{\kin}{\rm K1}
\newcommand{\kt}{\rm K05}
\newcommand{\tk}{\rm K01}
\newcommand{\cf}{CF}
\newcommand{\kinnp}{\rm K1_{np}}
\newcommand{\ktnp}{\rm K05_{np}}
\newcommand{\tknp}{\rm K01_{np}}
\newcommand{\vsfl}{\rm VSF^{los}}
\newcommand{\vsft}{\rm VSF^{3D}}

\newcommand{\kpc}{\mathrm{kpc}}
\newcommand{\Gyr}{\mathrm{Gyr}}
\newcommand{\kms}{\mathrm{km/s}}

\begin{document} 

\title{On the impact of AGN feedback modes onto the turbulent properties of the multiphase ICM} 

\author{S.  Sotira\inst{1,2}\fnmsep\thanks{\email{stefano.sotira@unibo.it}}, F. Vazza\inst{1,3} and F. Brighenti\inst{1,4} }
\institute{Dipartimento di Fisica e Astronomia, Università di Bologna, Via Gobetti 92/3, 40121 Bologna, Italy
\and
INAF - Osservatorio di Astrofisica e Scienza dello Spazio di Bologna, Via Gobetti 93/3, 40129 Bologna, Italy
         \and
            Istituto di Radio Astronomia, INAF, Via Gobetti 101, 40121 Bologna, Italy  
            \and
            University of California Observatories/Lick Observatory, Department of Astronomy and Astrophysics, Santa Cruz, CA 95064, USA
        }

   \date{Received / Accepted}


\abstract    
{
The feedback from active galactic nuclei (AGN) plays a crucial role in regulating the thermodynamics and the dynamics of the intracluster medium (ICM). 
Studying the turbulent patterns of the hot and warm ionized phases may allow us to determine how these phases are involved in the AGN cycle and the amount of turbulent pressure generated by the latter. 

In this work, we use new simulations to study the turbulent motions created by different types of AGN feedback in a cool core cluster and predict the observable signatures with the latest X-ray telescopes (e.g. XRISM).

We run several hydrodynamic simulations with ENZO, simulating the self-regulated cycles of AGN feedback, starting from a static ICM in a cluster that represents the Perseus cluster.
We study in detail different feedback modes: from pure kinetic precessing jets up to almost pure thermal feedback.

Our analysis reveals that the gas velocity dispersion in the center of the cluster correlates in time with the peaks of the AGN activity and that more than $50\%$ of the time, different feedback modalities produce the velocity dispersion observed in the Perseus cluster while leading to distinct geometrical distributions and velocity dispersion profiles.   
Moreover, we do not find a significant kinematic coupling between the hot and the cold phase kinematics.
We find a correlation between the AGN activity and the steepening of the velocity function structure (VSF)
and that the projected 2D VSF slopes are never trivially correlated with the 3D VSF ones.

This line of research will allow us to use incoming detections of gas turbulent motions detectable by XRISM (or future instruments) to better constrain the duty cycle, energetics and energy dissipation modalities of AGN feedback in massive clusters of galaxies.
}


\authorrunning{S. Sotira et al.}

\maketitle

\section{Introduction }
\label{sec:introduction}

Cool-core clusters of galaxies are objects which are believed not to have undergone major merger events in the recent past. Their intracluster medium (ICM) in the central region is characterized by a positive temperature gradient and a relatively short cooling time $\lesssim $ a few Gyr. 
The gas entropy profiles increase toward the external regions, and show a low-entropy core with values around $10-50\ \mathrm{KeV\ cm^2}$  \citep[e.g.][]{Fabian94,2010A&A...513A..37H,Sanders10}.

The (thermo)dynamics of these objects is believed to be self-regulated by the interplay of radiative gas cooling (causing overall inwards gas motions of the external gas layers around the core), and feedback events by the central active galactic nucleus (AGN) located in the central galaxy (or galaxies) at the bottom of the cluster potential well, triggered by the accretion of matter onto the central supermassive black hole (SMBH)\citep[for a review see ][]{Fabian12}. 
The ICM description still poses formidable open problems to theory that can be constrained knowing the turbulent structure of the gas: the hydrostatic bias, the masses obtained considering a hydrostatic equilibrium are underestimated with respect to the ones computed by the gravitational lensing, suggesting the presence of others contributions to the pressure support, beyond the thermal pressure \citep{Eckert19}; the metals distribution, i.e. how the turbulence diffuses and affects the metals profile in the ICM\citep{Simionescu08,Werner10}; the thermal heating due to the turbulence dissipation of the central AGN\citep{YangRey16,SijBou21,Zhuravleva14}.
In a cool core cluster, the sources of turbulence in the ICM can be several: cluster galaxy motions throughout the cluster; cosmological accretion or minor mergers occurring at the borders in the external regions \citep{va11turbo}; 
sloshing and the AGN activity in the central regions \citep{va12filter,heinz10,Randall15,YangRey16,Lau17,BourneSij17,BournSij19}.

The AGN activity of the central SMBH is certainly a key process that affect the cluster in many ways.
First of all, as mentioned above, the AGN feedback is thought to be the engine that regulates the thermodynamics of the cluster, heating the ICM but maintaining the cool core in a self-regulated way.
As suggested,  the feeding of the AGN is due to the cold gas that reaches the proximity of the SMBH \citep{Pizzolato2005,Pizzolato2010,Soker2006,gaspari12,ga13,werner14,Voit2015}.
The cold gas is observed at the center of galaxy clusters appearing as atomic and molecular phases \citep{Conselice2001,babyk2019,Olivares2019,Gingras24}.
It extends in regions of typical size of $10-70\ \kpc$ for the atomic phase and $10-20\ \kpc$ for the molecular phase.
It often appears with filamentary and clumpy shapes \citep{Olivares2019}.
The size of the clumps is likely smaller than $\sim 70\ \mathrm{pc}$, since the clumps are not resolved even with the highest resolutions \citep{Conselice2001,Olivares2019}.
The link between the AGN activity and the cold gas formation has been debated and is still not totally clear, but theories and simulations have made significant progress on this topic \citep{McNamara16,Voit2017,Gaspari18}.
The jets generated by the AGN are thought to be both heat sources and catalyzators of the cold gas formation: as the jets move through the ICM, they entrain hot gas from the center and the turbulence present at the edges of the jets helps the mixing of the layers and makes the cooling time shorter, favoring the condensation of the gas in clumps.
Then moved by gravity and losing angular momentum through several processes, the clumps fall toward the cluster center and the SMBH.
This process has been reproduced in several simulations \citep{gaspari11a, gaspari11b,LiCC2014,LiB15,Wang2021}.
The study the hot and cold gas motion and the link between them helps us to understand deeper the feeding and the heating of the AGN cycle.

The velocity structure function (VSF) is a tool to analyse the properties of turbulence.
VSFs of the cold phase have been calculated \citep[e.g.][]{li20,
Li23,Ganguly23,Gingras24} , while, because of the limitations of the current X-ray telescopes, VSFs of the hot phase are missing, except for some extraordinary attempts like \citet{gatuzz23}.
A great improvement in the hot gas kinematics observations is expected by the incoming observations of the XRISM telescope.

Simulations of kinetic mode feedback with different implementations have shown that this feedback mode is able to produce a level of turbulence observed at cluster center\citep{BourneSij17}. 
However, cosmological simulations aimed to study the turbulence throughout the ICM, considering also merging and cosmological accretions, usually use thermal AGN feedback since the resolution is not high enough for the implementation of kinetics jets.
Thermal feedback is also widely used because of its simplicity, even though the presence of kinetic jets is required by observations. It is not clear if purely thermal feedback can capture the basic physical processes at work in cluster centers.

Observations also suggest that precession of radio jets might be common \citep[e.g.][]{Martin13,Uber24}. 
The precession of the SMBH can be due to the lense-thirring effect or minor and major mergers of the clusters that end up with the merging of the corresponding SMBHs.
Also, accretion of stars or other episodes of large accretion has been proposed as origin of radio jets precession.

In this paper, we use hydrodynamic simulations of self-regulated AGN heating in cool-core galaxy clusters with different combinations of kinetic and thermal feedback, with and without a precession effect, to study the turbulence created by the different configurations and to check how the hot and cold gas kinematics are affected by these characteristics.
By Analysing the gas kinematics through bulk motion and velocity dispersion and the turbulent properties through the VSF allow us to constrain the physical properties of the AGN feedback, its efficiency in creating cold gas and the interactions between this phase and the hot ICM. 
With the upcoming insight due to the new generation of X-ray telescopes, we can be able to compare the simulation predictions with the real data.

The paper is developed as follows, in section \ref{sec:methods_sim} we explain the initial conditions and the setup of the simulation that is based on \citet{LiCC2014} and show the runs performed, in sec \ref{sec:therm} we have a glimpse about the thermodynamic of evolution to check if the AGN feedback works as cluster engine, in sec \ref{sec:kin} we show in detail the kinematics behaviour of the hot and the cold phases and in sec \ref{sec:vsf} we analyse the properties of the turbulence through the use of the VSF.
Finally, in sec \ref{sec:conclusions} we discuss the results obtained and give some future prospects.

\section{Methods and Simulations}  \label{sec:methods_sim}

In this paper, we adopt the simulation setup and feedback mechanism by \citet{LiB12} and \citet{LiB15}. 
Similar procedures have been used in several numerical studies, i.e.  \citet{gaspari11a,gaspari11b,BourneSij17,Wang2021},  and can be considered a state-of-the-art implementation of AGN feedback in single object simulations. Nevertheless, slight differences in the numerical parameters can generate very diverse flows and one of the main objectives of this work is to evaluate the sensitivity of hot and cold gas turbulence on the details of the feedback process. We will explore other feedback techniques in future works.

\subsection{Initial conditions}
We run our simulations with the ENZO code (enzo-project.org), a cosmological Eulerian code supporting Adaptive mesh refinement, and a large variety of different numerical solvers and physical modules \citep[][]{enzo14}.
Our simulations start from the numerical setup implemented in a series of works by Li \& Bryan and collaborators \citep{LiB12,Libryan14b,LiCC2014,LiB15}, which represents a cool core cluster with physical properties based on the Perseus cluster.

We summarise here the fundamental aspects of this model:

\begin{itemize}

\item The static gravitational potential follows the model by \citet{Mathews06} and is due to an NFW \citep{NA96.1} dark matter halo and a central Brightest Cluster Galaxy (BCG), with the following mass distributions: 
\begin{equation}
M_{\rm NFW}(y)=M_{\rm vir} \frac{\log(1+y)-y/(1+y)}{\log(1+c)-c/(1+c)}
\end{equation}
where $y=cr/r_{\rm vir}$, $c=6.81$
is the concentration, $r_{\rm vir} =2.440$ Mpc is the virial radius and $M_{\rm vir} = 8.5\times 10^{14}$ M$_\odot$.

The BCG stellar mass distribution is approximated with a fit of a De Vaucouleurs profile with a total mass of $2.43\times 10^{11}$ M$_\odot$ and effective radius $R_e = 6.41\ \kpc$:
\begin{equation}
    M_*(r) =\frac{r^2}{G}\Bigg[\Bigg(\frac{r^{0.5975}}{3.206\times 10^{-7}} \Bigg)^s +\Bigg(\frac{r^{1.849}}{1.861\times 10^{-6}} \Bigg)^s \Bigg]^{-1/s} ,
\end{equation}
in cgs unit with $s=0.9$ and $r$ in $ \kpc$
and a central supermassive black hole of $M_{\rm SMBH} = 3.4\times10^8 \msol$.

\item The initial gas temperature and electron density profiles are fits of X-ray observations of the Perseus cluster \citep{Churazov04,Mathews06}:

\begin{equation}
    T(r)=7\frac{1+(r/71)^3}{2.3+(r/71)^3} \ \rm keV,
\end{equation} 
\begin{equation}
n_e(r) = \frac{0.0192}{1+\big(\frac{r}{18}\big)^3}+\frac{0.046}{\bigg[1+\big(\frac{r}{57}\big)^2\bigg]^{1.8}}+\frac{0.0048}{\bigg[1+\big(\frac{r}{200}\big)^2\bigg]^{1.1}}\ \rm cm^{-3}.    
\end{equation}

These initial conditions represent an ICM in approximate hydrostatic equilibrium in the mass model described above. Therefore, the initial gas velocity is set to zero.

\item The gas loses energy through the cooling function based on table 4 from \citet{Schure09}.
A temperature floor of $T=10^4\ \mathrm{K}$ is set.

\item The simulations start with a root grid of side 64 cells in a comoving box of side $L=16\ \rm Mpc$, that is redefined by the AMR following criteria in order to resolve denser region, the cooling time and gravitational instability. 
In particular, a cell is refined when: the gas mass in the cell exceeds 0.2 times the gas mass in a root cell; the cooling time becomes small compared to the crossing time over the cell, in particular when $t_{\mathrm{cool}}<6t_{\mathrm{cross}}$; the size of the cell is larger than $0.25$ of the Jeans length.
\end{itemize}
More details can be found in \citet{LiB12}.

By using up to 10 levels of mesh refinement, the highest spatial resolution achieved in our simulations is  $\approx 240~\ \rm pc$.

\subsection{Accretion and feedback physics}\label{fb_implementation}
As stated above, we use a standard implementation of self-regulated AGN feedback.  
It is built on the long standing idea of cold accretion \citep[e.g][]{pizzoSoker05,BrighMath03,BrighMathew06} onto the central SMBH, which in turn triggers the injection of some form of energy in the surrounding ICM. Both steps must be parametrized in the numerical simulations. 
This mechanism naturally induces a self-regulated feedback cycle, the features of which, however, critically depend on the numerical details of the implementation. 
This class of algorithms has been widely used, refined and investigated in the last two decades \citep[just to cite a few ][for a review see \citealt{BourneYang23}]{Cattaneo07,gaspari11a,gaspari11b,gaspari12,Libryan14b,LiCC2014,LiB15,YangRey16,Cielo17,BourneSij17,SijBou21}.

The energy injection depends on the accretion rate, which is controlled by the quantity of the cold gas near the SMBH.
More specifically, the gas is accreted when the following conditions are fulfilled:
\begin{itemize}
\item the gas has to be colder than a threshold temperature $T_{\rm thres}=3\times10^4\ \rm K$,
\item the gas must be within a cubical region of side $1\ \rm  kpc$ around the SMBH. This is called the {\it accretion region}.
\item the mass of cold gas in the accretion region must be larger than a threshold value, $M_{\rm thres}=10^7 \rm \msol$. 
\end{itemize}

When a quantity of cold gas $M_{\rm acc}$ satisfied these conditions, the accretion rate onto the SMBH is estimated as $\dot{M}=M_{\rm acc}/\tau$, where $\tau=5\ \rm Myr$ roughly represents the free fall time in the accretion region.

The total power in the jet is 
\begin{equation}
    \Pjet = \epsilon \dot{M} c^2, \label{Agn_power}
\end{equation}
where $c$ is the light velocity and $\epsilon = 10^{-3}$ is the effective accretion efficiency.
To generate the jets, a quantity of mass (and energy, see below) is added  on two planes parallel to the equatorial plane, at a height on the $z$-axis $h_{\rm jet}= 2\ \rm kpc$.
The amount $\Delta m$ added in each cell of the planes  is calculated so that $\Sigma\Delta m=\Dot{M}\Delta t$ and $\Delta m\propto \exp{-r^2/2r_{\rm jet}^2}$, with $r$ being the distance from the $z$-axis and $r_{\rm jet} = 1.5\ \rm kpc$.

Needless to say, in reality some fraction $\Dot{M}$ would accrete onto the black hole and disappear from the flow.
Thus, a simplifying assumption here is to neglect the black hole growth and assume that $\Dot{M}=\Dot{M}_{\rm jet}$, i.e. the entire accreted mass gas is deposited in the jet region.

The AGN energy (and power) can be injected either in kinetic or thermal form in the jet launching region.
This is regulated by the parameter $\fkin$, which sets the fraction of the total feedback power converted into jet kinetic power, $\pkin = \fkin \Pjet$. 
We will see that predominantly "thermal" feedback (small $\fkin$) or predominantly "kinetic" feedback (large $\fkin$) lead to quite different flows.
The launching velocity of jets thus depends on the kinetic fraction $\fkin$, $v_{jet}^2=2\fkin \epsilon c^2$, through the formula $\pkin=1/2\Dot{M}v_{\rm jet}^2$.
In the jet launching planes, the momentum of each cell is $v_{\rm jet}\Delta m$, and the final velocity of each cell is computed considering that the total momentum is conserved.
When there is no thermal feedback, in the jet launching bases where the gas mass is added, the temperature is lowered in order to keep the thermal energy of the cell constant.
A side effect of this choice is an artificial decrease of the local cooling time, which favours cold clumps formation, as we will show in the discussion.
When there is predominantly thermal feedback, energy is directly injected as thermal energy, by adding to each cell of the jet launching region the quantity $\frac{1-\fkin}{\fkin}\frac{1}{2}\Delta m v_{\rm jet}^2$.
In this paper, we explore a range of the parameter $\fkin$ to investigate its impact on the feedback modes.  
We also consider the possibility of jet precession along the $z$-axis.
The jet has an precession angle of $\theta_{\rm p}\approx9^{\circ}$, and a period of $\tau_p=10\ \mathrm{Myr}$. 

\subsection{Runs}
\begin{figure*}
\begin{center}
\includegraphics[width=\linewidth,height=\textheight,keepaspectratio]{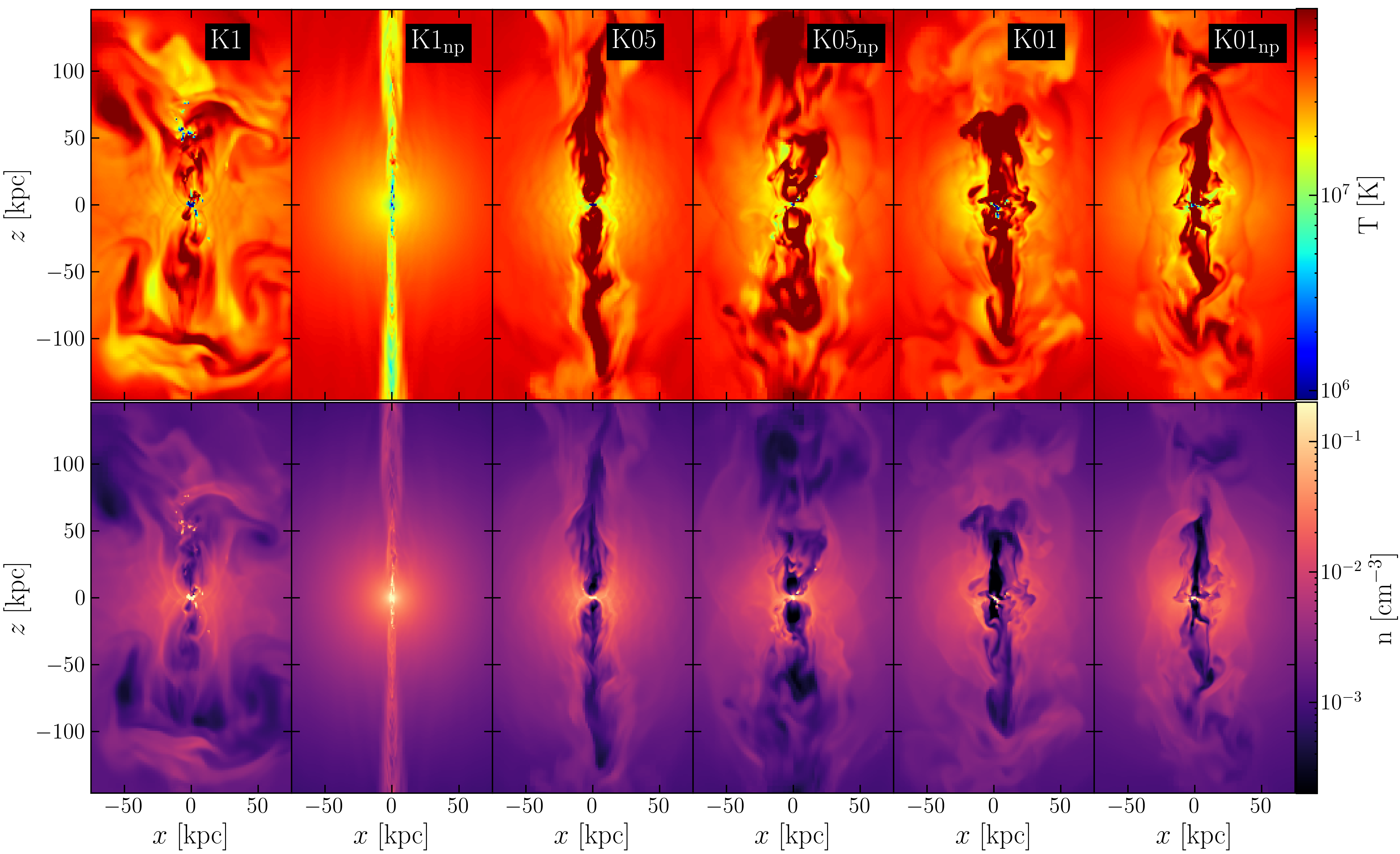}
\caption{Top panels: Gas temperature (top) and density (bottom) maps in the $x-z$ plane of $\kin$, $\kinnp$, $\kt$, $\ktnp$, $\tk$ and $\tknp$ runs  (from left to right). The images are not taken at the same absolute time, but rather at the epoch of the maximum AGN feedback power in each run. Movies for the precessing jets simulations can be watched at \url{https://www.youtube.com/watch?v=7HgSL4xY8lI} and at \url{https://www.youtube.com/watch?v=9_SnQm5R8Vs}.}\label{tem_dens_slice}
	\end{center}	
\end{figure*}
In order to investigate the differences in the feedback modes and the effect of jet precession we have performed six independent hydrodynamic simulations. 
In detail, the three feedback modes explored  here are: 
\begin{itemize}
    \item $\kinnp$, the feedback is purely kinetic ($\fkin =1$) and there is no jet precession. The model K1 is similar, but precession is present. 
    \item $\ktnp$, a mixed thermal and kinetic feedback modality, in which $\fkin=0.5$. Model K05  has also $\fkin = 0.5$ but includes jet precession.  
    \item $\tknp$, a predominantly thermal feedback model, $\fkin=0.1$, without precession. Similarly, model K01 has $\fkin=0.1$ and precession. 
\end{itemize}

We complement these feedback simulations with an extra run in which there is no feedback whatsoever,  \cf, with the feedback switched off for the entire evolution of the cluster and a classical cooling flow develops. 

In the next Section, we analyse the thermodynamic evolution of the ICM and the observable radial profiles in the various models, before analysing in detail the ICM kinematic.
Since the AGN cycle is self-regulated and the feedback modes are different, soon after the begin of the feedback stage we observe significantly different feedback and cooling evolution.

\section{Results: Thermodynamic evolution}\label{sec:therm}
Here, we present maps and profiles of the most relevant thermodynamical quantities. 
This analysis is straightforward but essential to establish if the simulated clusters agree with the basic X-ray observations. 
It is notoriously problematic, for heated flows, to satisfy at the same time the tight constraints on the cooling rate and having acceptable temperature and density profiles. 
This can be considered the first test that any realistic AGN feedback model must pass.
See, for instance, the discussions in \citet{BrigMath02,BrighMath03,BrighMathew06} and in \citet{gaspari11a,gaspari11b}

The six panels of Figure \ref{tem_dens_slice} show the temperature and density of the central slice around the plane $y=0$ of the six runs.
For each model, we display representative snapshots near the peak of activity of each AGN  model.
By simple visual inspection, a few salient features can be identified of our feedback implementations, also described below in a more quantitative way. 
The highest amount of perturbation in the ICM is observed in the purely kinetic feedback model $\kin$,  followed by the hybrid kinetic and thermal feedback without precession $\ktnp$ and finally by the two almost purely thermal with and without precession runs, $\tk$ and $\tknp$.
A distinctive and potentially observable feature is the persisting presence of very elongated cavities, or tunnels, along the jet direction. 
We also note (see Fig. \ref{cold_mass}) that in the purely kinetic run, the amount of the accreted or long living cold gas is larger than in $\ktnp$ run and the purely thermal cases $\tk$ and $\tknp$.
 This suggests  a close connection between the level of disturbance or turbulence of the hot ICM and the presence of inhomogeneities prone to thermal instability \citep[e.g.][]{gaspari12,Gaspari13}.
 
At variance with the previous models is run $\kt$, which forms a long lasting cold gas disk that leads the system to an accretion and feedback power almost constant with time, leaving the ICM less perturbed.
Finally, the purely kinetic run without precession $\kinnp$ soon runs into an unrealistic trend, in which most of the feedback energy is deposited at very large radii, incompatible with observations.  
We remark that this result is a consequence of the assumed accretion scheme, which generates a continuous accretion. This implies that the (not precessing and purely kinetic) jets mostly interact with a relatively smooth and unperturbed ICM, frequently generating new shocks and cavities. 
This underlines once more the impact of the numerical schemes on the simulated feedback physics.
It should be noticed that other schemes in the literature, like those used in \citet{gaspari11a,gaspari11b}, result in a more bursty AGN activity. 
 
In the following, we study the thermodynamic profiles of the ICM in order to test our models against X-ray observations, in a statistical sense.
In particular, we want to check which run is able to prevent the catastrophic cooling of gas in the cluster core and, at the same time, to maintain a quasi-steady balance between cooling and heating, preserving the cluster cool core.
We first analyse the different accretion and power histories of each run,  and compare their various thermodynamic radial profiles with X-ray observations of real clusters, in a similar mass range of our simulated Perseus cluster.
\begin{figure*}
\begin{center}
\includegraphics[scale=0.52]{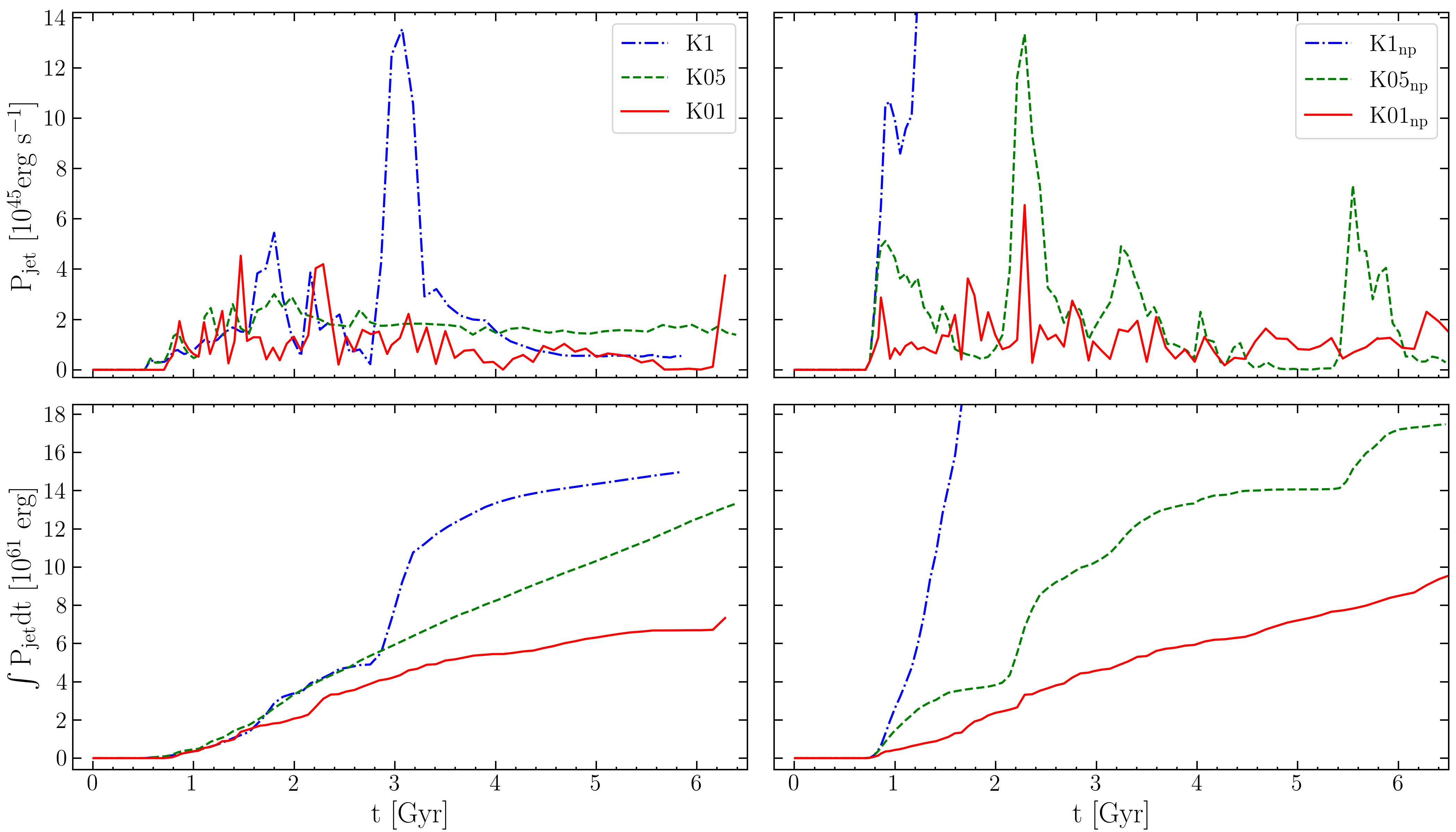}
\caption{Evolution of the jet feedback power (top) and integrated power (bottom) for the three runs with jet precession (right) K1 (blue, dotted-dashed), K05 (green, dashed) and K01 (red, solid), and without precession (left) with the same color and line-coding.}\label{power_jet}
	\end{center}	
\end{figure*}

Figure \ref{power_jet} shows the evolving feedback power recorded in all runs, as well as their total time-integrated feedback energy.
In our simulations, and as already found in many others \citep[e.g.][]{BrigMath02,Revaz08,gaspari11a,LiCC2014,Qiu20}, cold clumps often form in the outflows, out of the condensation of overdense material formed near the jet, and then fall back into the black hole. This circulation leads to a frequent feeding of the black hole over short periods. 
This can generate both strong, sparse AGN outbursts or a more continuous, smooth SMBH activity. Both regimes are allowed by the broad scenario of the Chaotic Cold Accretion (CCA) \citep[e.g.][]{gaspari11a,gaspari11b}. 

Different feedback modes heat the gas in the central region with different efficiency, and the additional presence of jet precession further affects the net heating of the cluster core, as well as the long term history of the injected feedback power. 
In particular, runs $\kin$ and $\ktnp$ have similar accretion histories up to $t \sim 3$ Gyr, when they experience the peak power of $1.5\times10^{46}\rm erg/s$. However, while in model K1 the AGN drops smoothly afterward, in model K05$_{\rm np}$ the AGN activity proceeds with a bursty regime, with several power peaks of about $4\times 10^{45}\ \rm erg/s$.
The latter behaviour is explained by the absence of precession, which facilitates the arrival of cold gas in the accretion region, while the former is due to the higher efficiency in creating dense cold gas clumps. 
Run $\kt$ has a lower and remarkably smoother power history. In this case, the formation of a cold disk leads to an almost constant accretion rate and a steady feedback power.
Runs $\tknp$ and $\tk$ show very similar trends, as expected, considering that the only difference is in the precession of kinetic jets, which however only account for a $10 \%$ of the total feedback power. 
They both have a lower, albeit irregular, history of feedback power, resulting in a total feedback energy lower than in the other cases, i.e. $6\times10^{61}\ \rm erg$.

The time integrated power for all the simulated models does not exhibit large differences, being limited in the range $7 - 15 \times 10^{61}$ erg (at $t=6$ Gyr), with the thermal models K01 and K01$_{\rm np}$ being a factor of 2 less energetic than the kinetic runs. The only exception is model K1$_{\rm np}$, which produces two thin and continuous jets which are unable to heat the gas in the cluster core. The jet power and energy quickly reach extreme values, but the heating is released at $\sim 1$ Mpc from the center. 
This peculiar evolution, which sometimes happens for purely kinetic, non-precessing narrow jets, causes a very non-isotropic -- and so not efficient -- ICM heating \citep[i.e.][]{VernReyn06}. 
As a counterexamples, the jets in \citet{BrighMathew06} or in \citet{gaspari11a,gaspari11b}, triggered in a more intermittent way, are able to distribute the heat in a wide volume.

\begin{figure*}
\begin{center}
\includegraphics[scale=0.6]{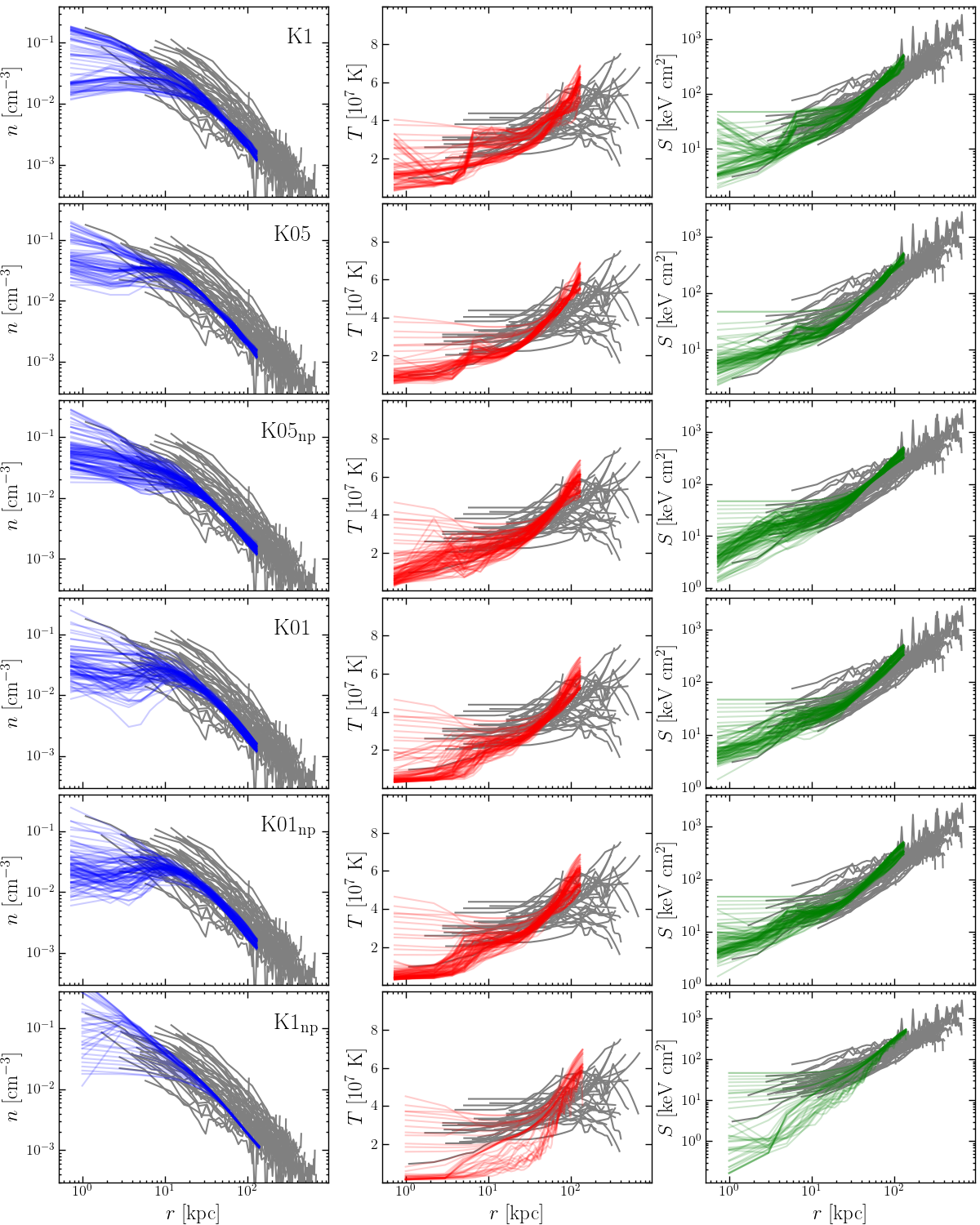}
\caption{Gas density (left panels), X-ray emission weighted temperature (center panels) and entropy radial profiles (left panels) obtained at several epochs during the cluster evolution in runs $\kin$, $\kt$, $\ktnp$, $\tk$, $\tknp$ and $\kinnp$ (from top to the bottom). The additional grey lines show the distribution of corresponding profiles obtained from the ACCEPT sample by \citet{2009ApJS..182...12C}, for clusters with a central temperature in a range of $3 \leq T \leq 5\ \mathrm{KeV} $ and a central entropy less than $50\rm \mathrm{KeV}/ cm^2$ meant to select cool core cluster with masses close to the Perseus-like cluster simulated.}
\label{profile_cavagnolo}
	\end{center}	
\end{figure*}
We now compare the simulated radial temperature, density and entropy profiles with the observed profiles in cool core clusters within the mass range of our simulated Perseus cluster, taken from the X-ray ACCEPT sample by   \citet{2009ApJS..182...12C}. 
Figure \ref{profile_cavagnolo} shows the density, the emission weighted temperature and the derived entropy profiles of some snapshots of our runs, overplotted on the observed ones from   \citet{2009ApJS..182...12C}.
Only the region $r > 10$ kpc is meaningful, since the inner zone is largely affected by the numerical implementation of the feedback mechanism.
We can see that for most outputs the profiles of our runs are in the same range as values of the observed one, except for the run $\kinnp$ for which the temperature and density profiles fail in the central $100\ \kpc$ size region. Notably, in all models, the drop in the central temperature, a signature of a cool core cluster, is preserved.

To summarize, we conclude that all runs, excluding $\kinnp$, are able to maintain, at least at first order, a balance between cooling and heating without breaking the cool core temperature profile.
Therefore, we move to the kinematic analysis of the five survived models in the next Sections.

\section{Results: Kinematic ICM analysis}\label{sec:kin}

To study the kinematic and turbulent properties of the ICM, we split the analysis into two components:
\begin{itemize}
    \item the hot phase: all the gas emitting X-ray radiation, i.e. cells with a temperature above $T=10^6\ \mathrm{K}$.  For the 2D projections, to compare our data with observations we weigh the velocities by the emissivity of the cell.    The emission weighted $i-$component of the velocity is: 
    \begin{equation}
        v_{i,\rm{hot}} = 
        \frac{\Sigma_{j=0}^{N}v_{i,j}n_j^2
        \Lambda(T_j)}{\Sigma_{j=0}^{N}n  ^2
        \Lambda(T_j)}\ \ \ \ \ T>10^6\ \mathrm{K}.\label{vproj_hot}
    \end{equation}

where the integration is performed along the line-of-sight in the $i$-direction and $N$ is the number of cells along the LOS specified at each computation.
    
    \item the cold phase: all the cells with a temperature below $T =5\times 10^4\ \mathrm{K}$.    This gas phase is usually observed near the cluster center, through emission lines such as $\rm{H\alpha}$, [OII], etc. The projected velocities of cold gas are computed by taking a constant value of the cooling emission for the
    H$\alpha$ line (case B recombination, $T=10^4$ K) $\Lambda_{\rm{H\alpha}}=3.3\times10^{-25}\ [\rm cgs]$ \citep{AgnAgn06}.
    \begin{equation}
            v_{i,\rm{cold}} = \frac{\Sigma_{j=0}^{N}v_{i,j}n_j^2\Lambda_{\rm{H\alpha}}}{\Sigma_{j=0}^{N}n^2\Lambda_{\rm{H\alpha}}}\ \ \ \ \ T<5\times 10^4\ \mathrm{K}.\label{vproj_cold}
    \end{equation}
\end{itemize}
Just for simplicity, we assume that there is neither neutral or molecular gas nor dust in the cold phase, which is optically thin, even if this  assumption is unlikely to be realistic. 

In the following, we solve the two sums along the line of sight of our observations, for different volume selections which are specified from case to case.

\subsection{Hot phase }

\subsubsection{Kinematical maps and $\sigma$ profiles}

\begin{figure*}
\begin{center}
\includegraphics[scale = 0.69]{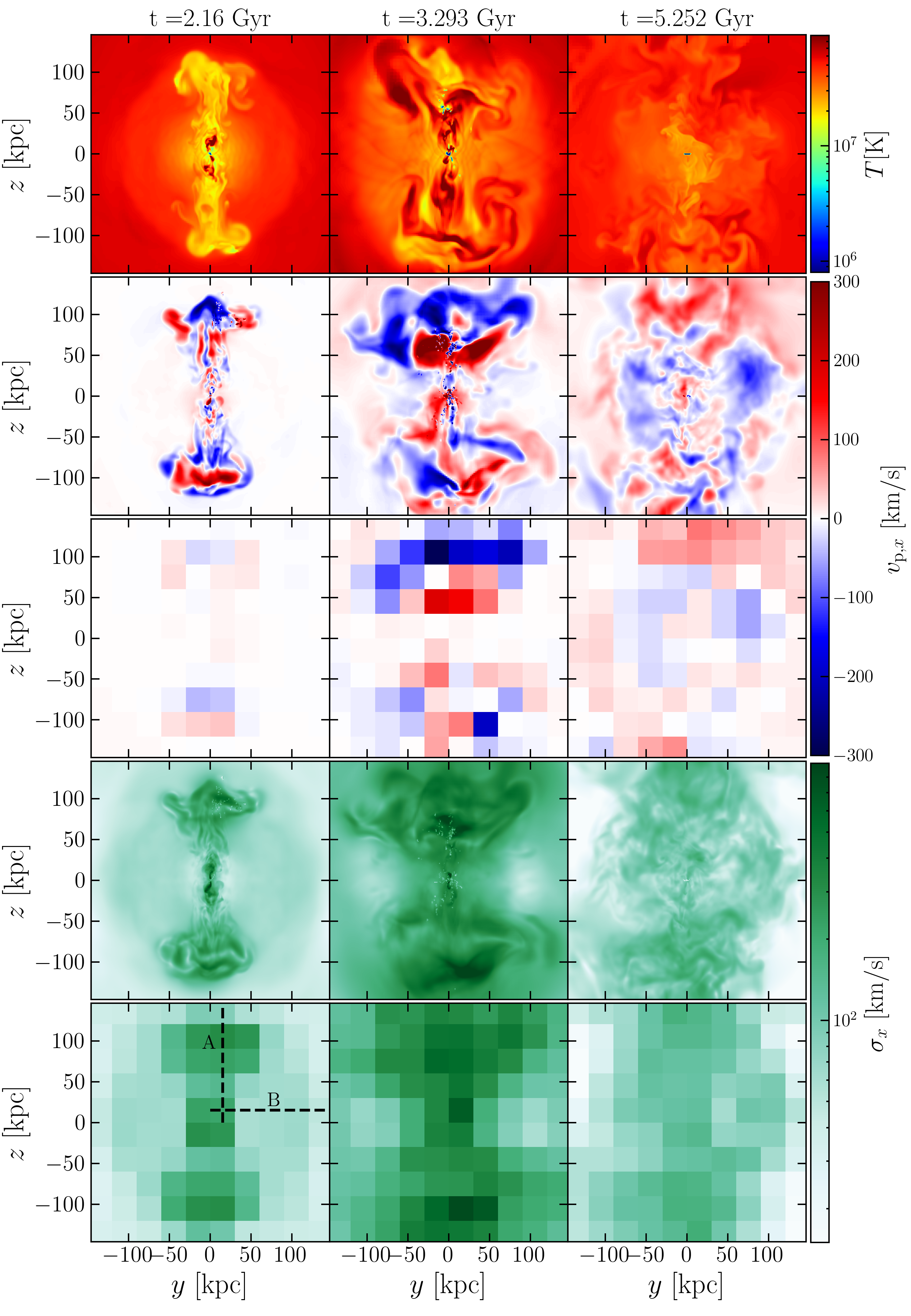}
\caption{
From left to right the quantities listed in the following at three epochs ($t =2.16,\ 3.29$ and $5.25\rm ~ Gyr$) of the kinetic run with precession $\kin$ are shown.
Top panels: temperature field in the meridional plane.
Second and fourth panels: X-ray emission weighted projected velocity map and corresponding velocity dispersion  obtained using eqs. \ref{vproj_hot} and \ref{eq_sigma_map} along a LOS perpendicular to the jet axis.
Third and fifth panels: grained X-ray emission weighted projected velocity and velocity dispersion maps obtained from Eq. \ref{v_pix} and \ref{sigma_pix} using a pixel of $30\times30\ \kpc$ along the same LOS perpendicular to the jet axis.
}\label{proj_velocity_x}
	\end{center}	
\end{figure*}
\begin{figure*}
\begin{center}
\includegraphics[scale = 0.69]{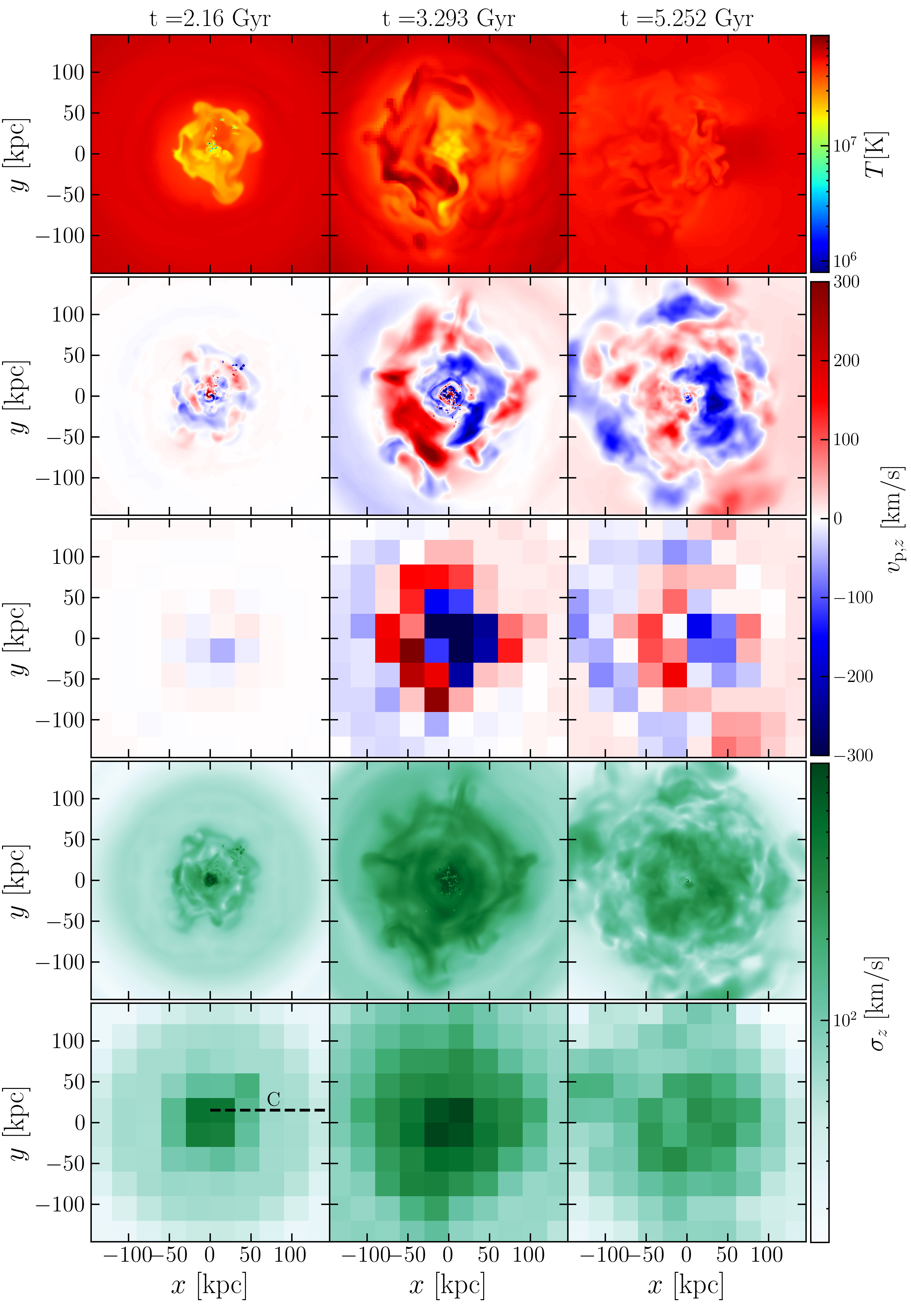}
\caption{
From left to right, we show several quantities  at three epochs ($t =2.16,\ 3.29$ and $5.25\rm ~ Gyr$) of the kinetic run with precession $\kin$.
Top panels: temperature field in a plane perpendicular to the jet axis at $100\ \kpc$ from the center.
Second and fourth panels: X-ray emission weighted projected velocity map and corresponding velocity dispersion  obtained using eqs. \ref{vproj_hot} and \ref{eq_sigma_map} along the LOS parallel to the jet axis.
Third and fifth panels: grained X-ray emission weighted projected velocity and velocity dispersion maps obtained from eqs. \ref{v_pix} and \ref{sigma_pix} using a pixel of $30\times30\ \kpc$ along the LOS parallel to the jet axis.}\label{proj_velocity_z}
	\end{center}	
\end{figure*}

\begin{figure*}
\begin{center}
\includegraphics[width=\linewidth,height=\textheight,keepaspectratio]{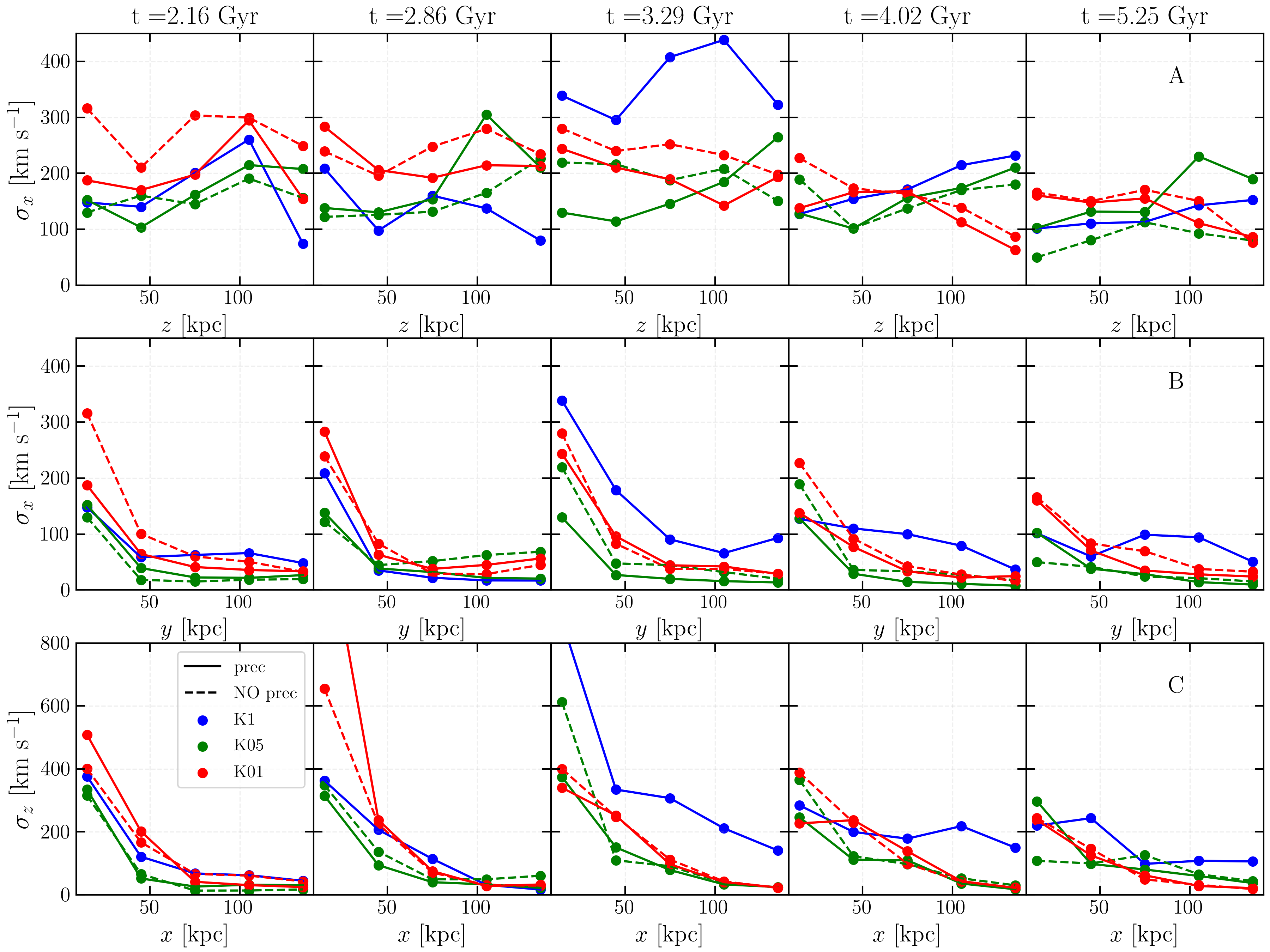}
\caption{Velocity dispersion profiles measured in pixels of $30\times30\ \kpc$ (Eq. \ref{sigma_pix}) taken along the $z$ axis (first row) and the $y$ axis (second row) for the map projected perpendicularly to the jet axis (as indicated by the black dashed lines A e B in Fig. \ref{proj_velocity_x}) and along the $x$ direction (third row) for the map projected parallel to the jet axis (as indicated by the black dashed line C in Fig. \ref{proj_velocity_z}) for the runs $\kin$ (blue solid ),$\kt$ (green solid), $\ktnp$(green dashed), $\tk$ (red solid) and $\tknp$ (red dashed) taken at various epochs $t=2.16, 2.86, 3.29, 4.02$ and $5.25\ \Gyr$ (from left to the right).} \label{fig:sigma_pro}
	\end{center}	
\end{figure*}

We start by studying the kinematics of the hot gas phase by calculating quantities observable (or potentially observable) in X-rays. We focus first on model K1 and then briefly discuss the other runs.
We recall that in our simulations the ICM is set initially completely at rest (i.e. no rotation or random motion is set in the initial condition).  At first, a classic cooling flow stage sets in, with timescale $ t_{\rm{cool}}\approx 1 \Gyr$, during which the gas starts flowing toward the center with average velocities of the order of $\simeq 10\ \kms$.
When the feedback gets activated, the gas along the jet axis begins to experience strong accelerations. 
The first two top rows of Figure \ref{proj_velocity_x}  show the 3D temperature in the $z-y$ plane (first panel) and the X-ray emission weighted projected velocity along the $x$-axis, a LOS perpendicular to the jets (second panel) at three different epochs of the run $\kin$, with the time increasing from left to the right.
In the same way, the first two rows of Fig. \ref{proj_velocity_z}
show a temperature map on a plane defined by $z=100\ \kpc$ and the X-ray emission weighted $z$-component of the velocity, $v_z$, projected along the $z$-axis (that is, the jet axis).

During the first episodes of AGN feedback ($t=2.16\ \rm Gyr$ in the images), the jets pierce the  ICM and start to expand laterally, at a height $z\simeq 100$ kpc. 
The jet material mixes with the ICM through dynamical instabilities at the jet edges, yet the motion of the  gas is mainly confined within the cone traced by the precession of jets.  The first velocity maps (first epoch of Figs. \ref{proj_velocity_x} and \ref{proj_velocity_z}) allow us to approximately trace this cone. 
At later times ($t=3.29\ \rm Gyr$), the AGN has produced several outbursts and we are close to the peak of the feedback power. 
Jets have inflated hot and underdense bubbles, which in turn set in motion the gas at a larger lateral distance from the jet cone (as shown in the second epoch of Figs. \ref{proj_velocity_x} and \ref{proj_velocity_z}).
Later on, after $\approx 4\ \mathrm{Gyr}$ since the first feedback event, gas motions are roughly isotropic and volume filling in the central cluster regions  ($t=5.25\ \rm Gyr$ in Figs. \ref{proj_velocity_x} and \ref{proj_velocity_z}).

As AGN feedback can be a key driver of turbulent motions in the core of clusters of galaxies \citep[e.g.][]{Gaspari18}, we also  computed maps of the gas velocity dispersion, projected along the line of sight.  
For example, the velocity dispersion along the $x$ direction is computed as 
\begin{equation}
    \sigma^2_{x}(y,z) =\frac{\Sigma_{i=0}^{N}(v_{x,i} - v_{p,x})^2\Lambda(T_i) n_i^2}{\Sigma_{i=0}^{N}\Lambda(T_i)n_i^2},\label{eq_sigma_map}
\end{equation}
where the sum is done along the $x$ direction, $v_{p,x}$ is the EW projected velocity in a given point $(y,z)$ on the projection plane, obtained with Eq. \ref{vproj_hot}
and $N$ is the number of cells along the LOS We use a cubic volume of side $300\ \kpc$ to perform the computation.
In a similar way, $\sigma_z$ is computed along the $z$ direction.
The fourth row of each epoch in Figs.  \ref{proj_velocity_x} and \ref{proj_velocity_z} show the map of the velocity dispersion along the $x$ and $z$ directions, for model K1.

The velocity dispersion is initially of the order of $10^2\ \kms$ within the jet cone, and only a few 10s of km/s outside the cone, typical of a (slow) cooling flow (left panel).
Later on, when the AGN activity is strong, the perturbed region widens, with $\sigma_x$ increasing up to $\sim 450 \;\rm km/s$ (central panel). After the strong outbursts and during the period of relatively lower AGN activity, the velocity dispersion stabilizes at values of few $\sim 10^2 \;\rm km/s$. Now $\sigma_x$ is more homogeneously distributed in the whole field considered for these projections ($300 \times 300 \;\rm kpc^2$), albeit the turbulent region maintains a broad cylindrical symmetry along the jet axis.

The turbulent region expands with time roughly with velocity $\approx \sigma_x \sim 100$ km/s, which we confound with the characteristic turbulent velocity, rather than the sound speed, about an order of magnitude larger.
The other runs have,  on average, a lower velocity dispersion (see also Fig. \ref{sigma_xyz_hotCold}), and the region in which the turbulence diffuses is correspondingly smaller.

Finally, the third and the fifth panels of Figs. \ref{proj_velocity_x} and \ref{proj_velocity_z} show the emission weighted projected velocity and velocity dispersion, as seen with a spatial resolution of $30\times 30\ \kpc$, for qualitative comparison with the observations of the Perseus cluster by the Hitomi satellite \citep[][]{hitomi16,2019SSRv..215...24S} and future observations with XRISM. 
For every pixel the projected velocity in that pixel, $v_{\rm pix}$, is computed using the following formula:
\begin{equation}
    v_{\rm pix}=\frac{\Sigma_{\rm cells} \Sigma_{i=0}^N v_{x,i}\Lambda(T_i)n_i^2}
    {\Sigma_{\rm cells}  \Sigma_{i=0}^N\Lambda(T_i)n_i^2},\label{v_pix}
\end{equation}
where the first sum is over the numerical cells in the $y-z$ plane (the ``plane of the sky'') within the pixel area. The second sum is over the cells along the $x$ direction (that is, it performs the integration along the line-of-sight).
Likewise,  the velocity dispersion $\sigma_{\rm pix}$ in each pixel is computed as
\begin{equation}
    \sigma_{\rm pix}=\frac{\Sigma_{\rm cells} \Sigma_{i=0}^N(v_{x,i}-v_{\rm pix})^2\Lambda(T_i)n_i^2}
    {\Sigma_{\rm cells} \Sigma_{i=0}^N\Lambda(T_i)n_i^2}.\label{sigma_pix}
\end{equation}
where $v_{\rm pix}$ is from Eq. \ref{v_pix}.
The sum along the LOS extends out to 300 kpc, to include the contribution of the foreground and background ICM.
It should be noticed that, although most of the information related to this X-ray weighted quantity comes from the dense cluster core, which is fully sampled by our procedure, we are neglecting a relatively small contribution from the outer layers in the cluster, due to computing limitations.

It is clear that strong AGN outbursts like the one at $t\sim 3$ Gyr for model K1 are easily detectable even at relatively coarse spatial resolution.
Both the velocity and the velocity dispersion achieve extreme values $\gtrsim 250$ km/s, just like in the true maps displayed in the second and fourth rows of Fig. \ref{proj_velocity_x} and \ref{proj_velocity_z}. 
The dilution induced by the decreased resolution do not erase this signature of violent feedback events.

Furthermore, the kinematic maps provide important information about the geometry of the outflows. When the jet axis is almost perpendicular to the LOS, both the velocity and dispersion display a bimodal distribution on the maps, with two highly perturbed regions symmetrically placed with respect to the cluster center (see Fig. \ref{proj_velocity_x}).
On the contrary, when the outflows are aligned with the LOS, only one central peak of velocity and dispersion is present (Fig. \ref{proj_velocity_z}).
During periods of weaker AGN activity, the maps are more homogeneous with average values for the velocity and dispersion reduced by a factor of $\approx 2$ with respect to the activity peak period.
To study the geometrical differences due to the LOS just discussed and to compare the velocity dispersion obtained in the other runs, we plot the velocity dispersion profiles taken along two perpendicular directions for the edge-on projection since the distribution appears axi-symmetric, and one for the face-on projection which is symmetric.  
The first two rows of Fig. \ref{fig:sigma_pro} show the velocity dispersion profiles of all the runs obtained from the grained maps projected perpendicularly to the jet axis along the $z$ and $y$ direction respectively (as indicated by the dashed black lines A e B of Fig. \ref{proj_velocity_x}).
The third row of Fig. \ref{fig:sigma_pro} shows the profiles obtained from the grained maps projected parallel to the jet axis along the $x$ direction (as indicated by the dashed black line C of Fig. \ref{proj_velocity_z}).
The velocity dispersion profiles are very different, depending on the considered regions as well as on the  feedback modality.
In particular, when the jets are observed face-on (third row of Fig. \ref{fig:sigma_pro}) there is a large drop in the $\sigma$ profile, especially when the AGN activity is strong.
The same behaviour, with a smaller jump, occurs when we observe the jets edge-on and move perpendicularly to the jet axis (second row of Fig. \ref{fig:sigma_pro}).
Conversely, when we see the jets edge-on but we move close to the jet axis, the velocity dispersion profiles show oscillating behaviours when the AGN is active and increasing or decreasing profiles, depending on the run, when the AGN is quiet (late epochs).
The shape of each profile is very dependent on the run, i.e. the feedback modality and the presence of precession. We conclude that, while single pointed measurements of the gas velocity dispersion in specific regions of the cluster are generally too affected by time variations to constrain feedback parameters, the full reconstruction of the profile of gas velocity dispersion across a few hundred of kiloparsecs has the potential to discriminate between different AGN feedback modes.

\subsubsection{Evolution of velocity dispersion}

The next step is the analysis of the time evolution of the central hot gas velocity dispersion. 
This is a key quantity to characterize the turbulence or bulk motion of the ICM.
\begin{figure*}
\begin{center}
\includegraphics[width=\linewidth,height=\textheight,keepaspectratio]{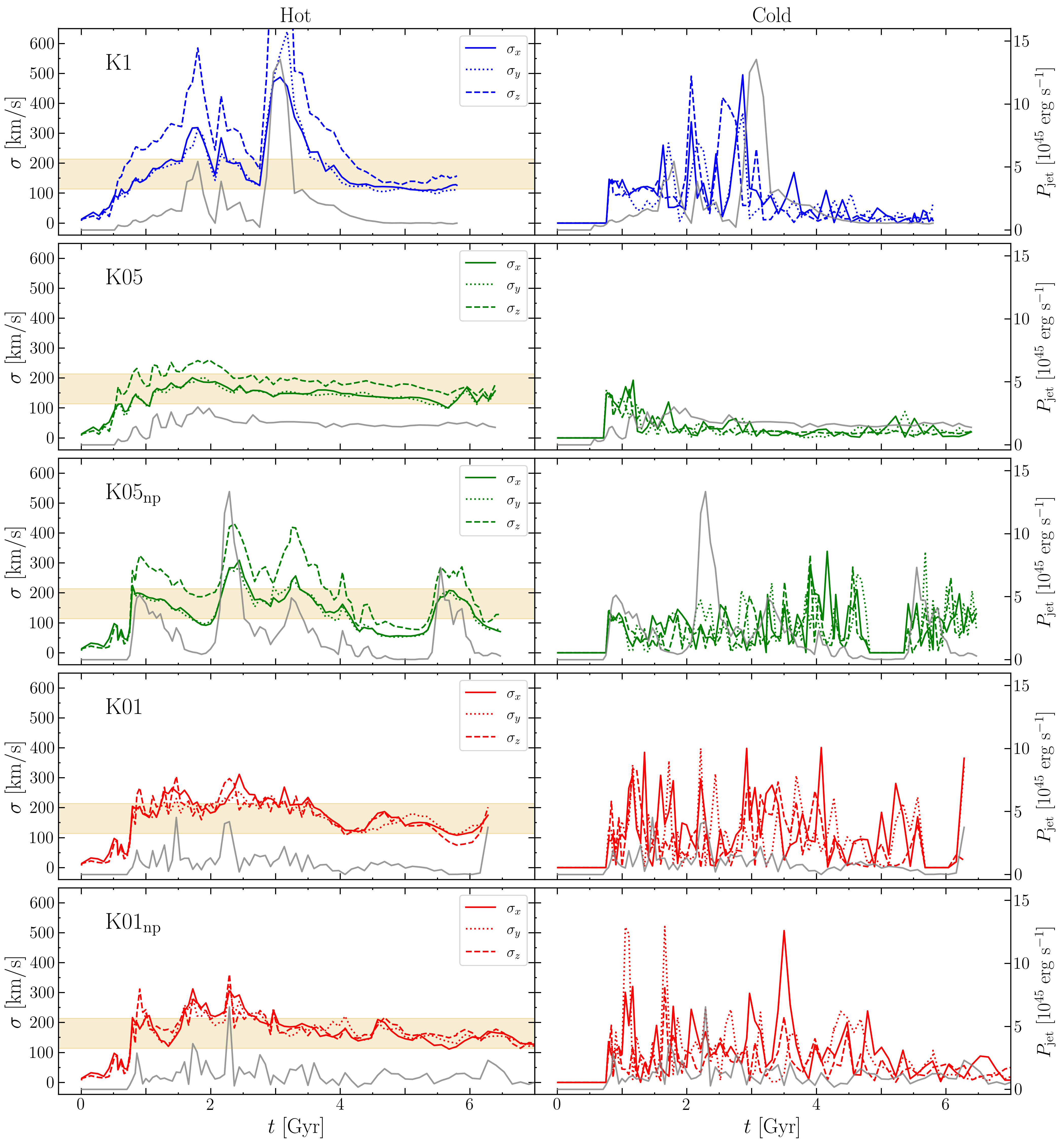}
\caption{Velocity dispersion of the hot gas (left) and of the cold gas phase (right) computed in a region of size $60\times60\ \mathrm{kpc}^2$ with an integration LOS of $150\ \kpc$, projected along the $x$ (solid), $y$ (dotted) and $z$ (dashed) axis, for the runs $\kin$, $\kt$, $\ktnp$, $\tk$ and $\tknp$ (from top to the bottom). 
The grey lines show the evolution of the feedback power at the corresponding times.}
\label{sigma_xyz_hotCold}
	\end{center}	
\end{figure*}

Since we are focusing on only the value of the central region of the cluster, we decide to consider a larger volume, compared to the pixel considered before, of size $60\times60\ \mathrm{kpc}^2$ in the projected plane, with an integration over a LOS of $150\ \kpc$.
The first column of Fig. \ref{sigma_xyz_hotCold} shows the three projections of the velocity dispersion $\sigma_x, \; \sigma_y$ and $\sigma_z$, of the hot phase versus time for the models $\kin$, $\kt$, $\ktnp$, $\tk$ and $\tknp$ (from top to the bottom). 
The velocity dispersion is computed similarly to Eq. \ref{eq_sigma_map}. 

The first thing to note is the strong time correlation between the AGN activity (light gray line) and the velocity dispersion behaviour, with the latter showing an increase right after the first powerful AGN burst, and following the later peaks of the AGN feedback during the whole evolution. The correlation is more evident in the runs with a more intermittent AGN activity and is of course expected considering that AGN feedback is here the most relevant driver of gas motions.

The velocity dispersion is strongly anisotropic in the $z$ direction for the run $\kin$ and it is also anisotropic for models $\kt$ and $\ktnp$, while is almost isotropic in the runs with a major thermal feedback mode, $\tk$ and $\tknp$.
Considering the velocity dispersion projected along the LOS perpendicular to the jet axis, which is less directly affected by the initial jet velocity, the run $\kin$ shows velocity dispersion values around $200 \kms$ for most of the time, while during the strong outbursts, it reaches values around $450\ \kms$. 
Contrary to K1, run K05 exhibits an almost constant velocity dispersion, with value $\sim 150$ km/s, with little
difference among the various projections\footnote{We remark that in these computations, the central velocity dispersion is lower than the central one measured in maps and profiles, due to the larger region considered here.}.
The run $\tk$ and $\tknp$ show $\sigma $ values around $200\ \kms$, with peaks reaching $300\ \kms$ after strong AGN activities.

It is important to compare the  velocity dispersion measure above with the actual measurement of the gas velocity dispersion obtained for the  central region of the Perseus cluster by the Hitomi satellite  \citep[][]{hitomi16,2019SSRv..215...24S}, which represents the best constraint on the  kinematics of the hot ICM at the center of the cluster to-date. 
Hitomi measured a relatively quiescent atmosphere, characterized by a gas velocity dispersion of $164\ \pm 15\ \kms$ in a region of  $60\times60\ \mathrm{kpc}^2$ around the core of Perseus.  
The horizontal golden band in the Figure marks the range of values $\sigma=164\pm50\ \kms$. 
The table \ref{tab:sigma}
reports the total span of time in which the velocity dispersion stays in these values for all the runs, during the first episode of feedback $\simeq0.8\ \Gyr$ and $6\ \Gyr$.
In general, the range of dispersion values measured by Hitomi is well reproduced by all our AGN models,  although in each different AGN modality, the chance of observing this specific value is different. 
The velocity dispersion in run $\kt$ remains in the observed range of values for almost the entire evolution, followed by the thermal runs, $\kt$ and $\ktnp$. 
For the $40\%$ of their evolution  (and strongly correlating with strong AGN bursts) run $\ktnp$ and, even more, run $\kin$ show an excess of gas velocity dispersion well beyond the Hitomi constraint. 

Combined with the results of the previous Section, this suggests that, while measurements taken in specific portions of the clusters and at specific times can hardly discriminate among different AGN feedback scenarios, the combination of spatially distributed measurements of gas velocity dispersion, and possibly the complementary measurement of cold gas velocity statistics (see next Section) better constrain the actual modality of deposition of feedback energy from AGN.

\begin{table}
    \centering
    \begin{tabular}{cccccc}
    \hline
LOS  &$\kin (\%) $ & $\kt$ &$\ktnp$& $\tk$& $\tknp$ \\
    \hline
$x$ & 63 & 89  & 52  & 64 & 64  \\ 
$y$ & 54 & 84  & 44  & 73 & 67  \\
$z$ & 43 & 71  & 18  & 53 & 64   \\
    \hline
    \end{tabular}
    \caption{Percentage of time during which the velocity dispersion computed in the central region of size $60\ \kpc$ stays on the values $164\pm50\ \kms$, during $5.2\ \Gyr$, from the first feedback event.}
    \label{tab:sigma}
\end{table}

\subsection{Cold phase}
\begin{figure*}
\begin{center}
\includegraphics[width=\linewidth,height=\textheight,keepaspectratio]{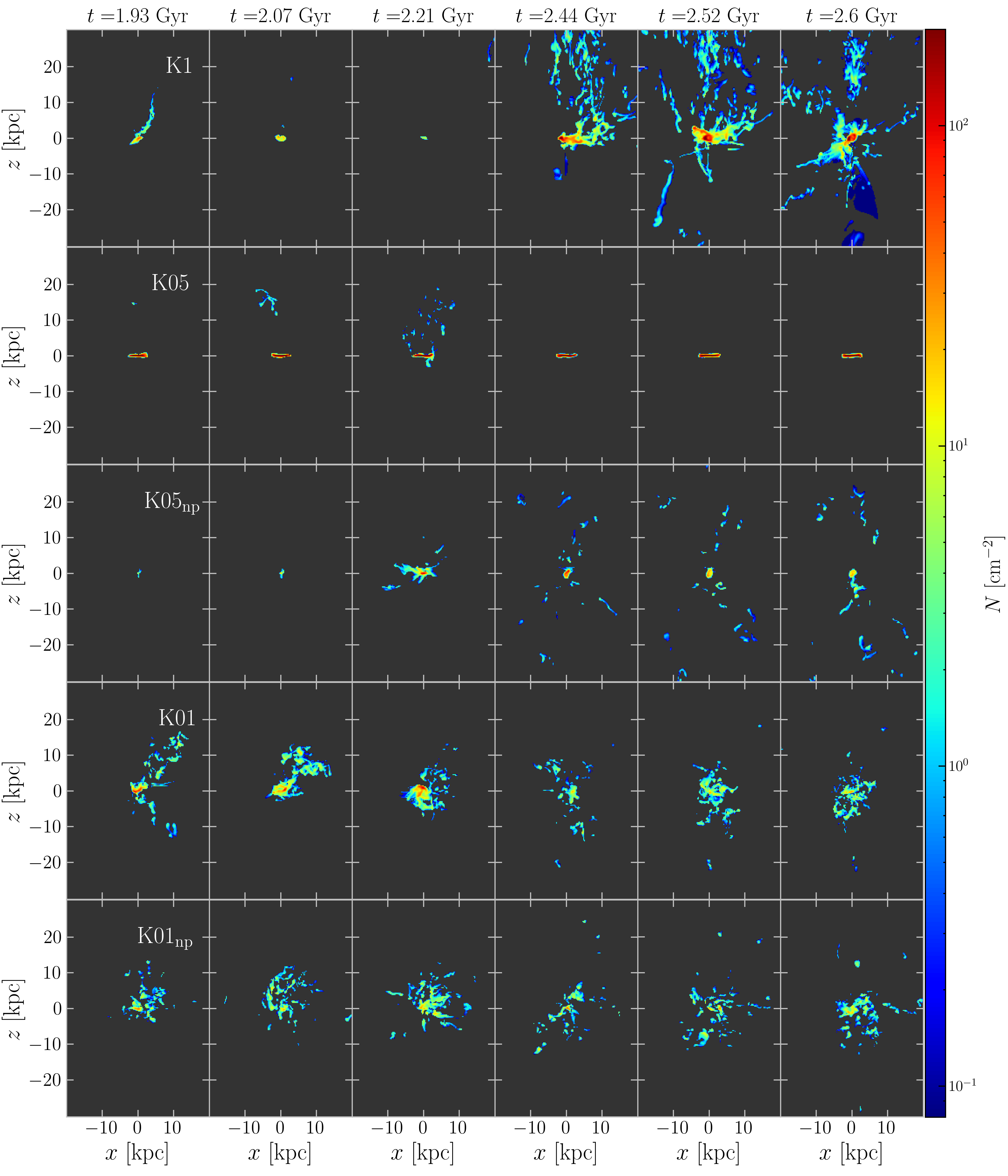}
\caption{Column density of the cold gas,$T<5\times10^4\ \mathrm{K}$, projected on a plane parallel to the jet axis, for various snapshots ranged in less than $1\ \Gyr$ (from left to the right), of the five runs $\kin$, $\kt$, $\ktnp$, $\tk$ and $\tknp$ (from top to the bottom). The figure shows the crucial differences in the cold gas formation: the run $\kin$ shows the most prominent cold gas structures in a filamentary and clumpy form. The run $\kt$ shows the presence of a long living cold gas disk, with very rare formation events of few clumps.
While run $\ktnp$ forms less cold gas compared to runs with precession, it shows a spatially large distribution of cold structures, similar to run $\kin$. Runs $\kt$ and $\ktnp$ have cold gas structures more localized in the center, with the former, with precession, having more cold gas.}\label{cold_gas_maps}
	\end{center}	
\end{figure*}
\begin{figure*}
\begin{center}
\includegraphics[width=\linewidth,height=\textheight,keepaspectratio]{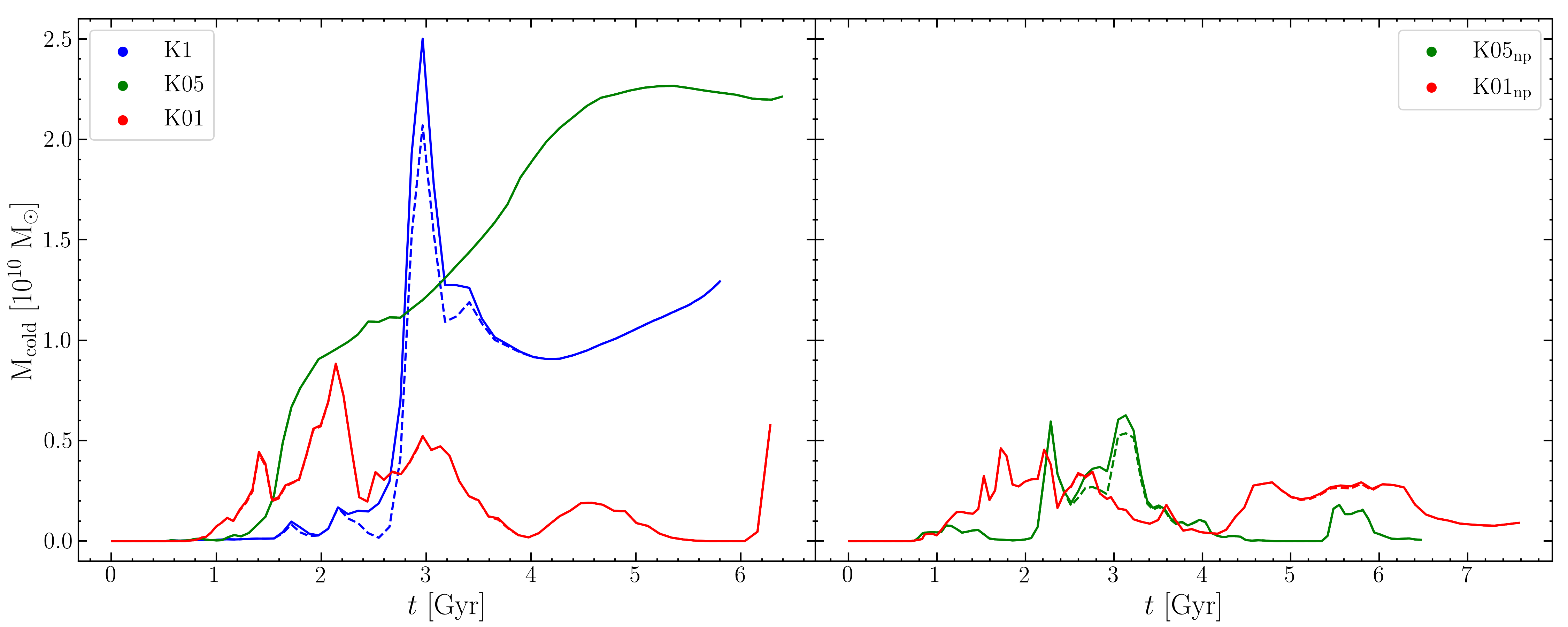}
\caption{Amount of gold gas mass at each snapshot, and not accreted in the SMBH (i.e. remaining outside the accretion region) computed in volumes of side $150\ \kpc$ (solid line) and of side $50\ \kpc$ (dashed line) for the runs with precession $\kin$, $\kt$ and $\tk$ (right) and without precession $\ktnp$ and $\tknp$ (left). For the runs $\kt$, $\tk$ and $\tknp$ the lines overlap.}\label{cold_mass}
	\end{center}	
\end{figure*}

Next, we analyse in detail  the properties of the cold gas, defined as a gas phase with $T<5\times10^4\ \rm K$.
In our simulations, cold gas forms by radiative cooling the hot ICM, with an efficiency that critically depends on the interplay between feedback and precession (or lack thereof).
Particularly interesting is the gas which cools off-center, usually at a distance of a few 10s kpc.
Such spatially distributed cooling, resulting by sustained, localized compression of the ICM and/or lifting of low entropy gas originally located near the BCG, is thought to play a key role in the feedback process 
\citep[e.g.][]{Gaspari13,Gaspari20,Voit2017}. The physical conditions that lead to off-center cooling have been investigated by many authors\citep{BrigMath02, gaspari11a,gaspari11b,gaspari12,mccourt12,Sharma13,LiCC2014,Libryan14b,valentiniBrigh15,BourneSij17,Qiu20} who tried to identify the main physical source of the (non-linear) perturbations needed to start the cooling process. 
Either bulk motion, powered by AGN outflows or cavity buoyancy, turbulence or lifting -- more likely a combination of these -- can do the trick (see references above). 
All these processes are effective if and when the susceptibility of the ICM to cool down is high. 
This property can be quantified through the cooling time $t_{\mathrm{cool}}$, the cooling-to-dynamical time ratio, $t_{\mathrm{cool}}/t_{\mathrm{dyn}}$, the entropy parameter $T/n^{2/3}$, and so on.

The study of the cold gas kinematics is crucial. 
First, it constrains the origin of cold gas \citep[e.g.][]{Gaspari18} and the nature of AGN feedback. 
Second, cold (ionized and molecular) gas can be observed in great detail with current instruments \citep{li20,Hu22,Ganguly23,Gingras24}, and this wealth of data represents a crucial test bed for simulations \citep[see][]{Wang2021}.

In most of our models, the combination of the parameters of the (central) ICM and the perturbations induced by the feedback generates cold gas in an intermittent fashion. 
We generally find two main types of structures: a population of blob-like cold clumps, often arranged in radial filaments (Figure \ref{cold_gas_maps}),  and a central cold rotating disk around the SMBH.
In the  following, we study several kinematic quantities, like velocity dispersion and the first order velocity structure function, to characterize the motion of cold gas and its relation with the feedback models.

\subsubsection{Cold disk}
In a few runs (late epochs of $\kin$ and most of the time of $\kt$) a long living rotating cold gas disk, with masses around $10^{10}\ \msol$, forms in the very central region of the cluster.
Unlike for the case of the cold clumps (see later), we think that the fate of long living rotating disks in our simulations is entirely (or mostly) due to the lack of physical ingredients as well as to specific way in which AGN feedback is implemented here, and therefore our predictive power in this region is extremely small. 
In such high density regions, the physical processes that are expected to govern the long term evolution of the cold gas, like star formation and its consequent feedback are missing in our simulations.
There is no radiative feedback from the AGN that is expected to be impactful especially very close to the SMBH.
Moreover, all the cold gas chemistry below the set temperature floor of $10^4\ \mathrm{K}$ is missing.
Indeed using the same setup, but including star formation \citet{libryan15} showed that the presence of massive rotating structures is inhibited by the onset of star formation.

Another factor that makes it hard to assess the realism of the forming cold gas is how  feedback is numerically implemented here: the jets start in two disks at a certain height from the center and this configuration tends to leave  the gas in the center unperturbed, promoting the formation of angular momentum along the vertical axis. 
Moreover, the formation of angular momentum along a coordinate axis, starting from gas completely at rest, is often found in fixed grid simulations.   Conversely, transient events of rotating cold gas that do not end in massive discs should be more realistic.

\subsubsection{Cold clumps}

Frequently observed structures in our runs are small cold clumps, as also found in \citet{LiCC2014} and papers quoted above.
These clumps are under-resolved in our simulations, with a typical size of 2 -- 10 cells (i.e. $0.5-2.5\ \kpc$), often with quasi-spherical shape. This implies that the physical evolution of cold gas cannot be accurately followed.
The clumps can form filament-like structures (see models $\kin$ and $\ktnp$ in Figure \ref{cold_gas_maps}) or
they can physically merge, especially in the central region, forming larger blobs. Previous works \citep[e.g.][]{Wang2021,Ehlert23,DasGronke24} have suggested that in the presence of a significant ICM magnetic field, the cold gas naturally assumes a filamentary shape.

A caveat in the cold clumps formation or cold gas in general due to the numerical implementation has to be discussed.
As described in sec. \ref{fb_implementation}, new gas mass is added at the jet launching regions, followed by a proportional decrease of gas temperature in order to keep the thermal pressure constant.
Thus, the adopted implementation artificially reduces the cooling time of these cells ($t_{\rm cool} \propto \sqrt{T}/\rho$ in the case of bremsstrahlung losses). 
It follows that in the outflowing jet material, the formation of thermal instability is numerically accelerated.
This potentially important effect might be present in other implementations of AGN feedback and the quantitative impact on the formation of cold clumps will be studied in a future paper.

Nevertheless, the condensation of the cold gas in clumpy or filamentary structures is seen in almost all simulations presented here and in the recent literature. 
Several general properties of cold gas nebulae (size, location, basic kinematics) agree with observed cool core clusters \citep[e.g.][]{Gaspari18}, so despite the aforementioned numerical process, which may be acting like a "catalyst", the formation of cold gas is  a real physical process.
The mechanisms which originate the so called ``precipitation'' process are several, from the entrainment by the jets/cavities of low-entropy gas living in or near the cD galaxy to the generation of non-linear perturbations by (subsonic) turbulence or bulk motion     \citep[e.g][]{BrigMath02,Revaz08, gaspari11a,LiCC2014,Brigh15,valentiniBrigh15,mccourt12,Voit2017}. 
Currently, it is not clear which process dominates in real clusters.

\subsubsection{Cold gas formation}
In the following, we will focus on the cold gas structures formed in the different runs.
We will focus only on that phase of cold gas that it is {\it not} accreted at each specific moment onto the SMBH, which represents the potentially observable cold gas mass at each epoch. 
In reality, the gas that reaches the very center of the cluster is either accreted onto the SMBH or experiences other processes\footnote{Actually as explained in Sec. \ref{sec:methods_sim}, all the cold gas reaching the center region of radius $500\ \mathrm{pc}$ is removed, but the part that goes in AGN power is only $1\%$ of this mass, this is an attempt of taking in consideration all the difficulties of the gas in reaching effectively the SMBH, as heating processes, the accretion disk and so on.}, but our study aims to analyse the cold gas around the center and to compare its properties to real observations. 
Of course, if this gas does not get destroyed by heating processes, it will eventually fall into the accretion region so into the SMBH, unless it gets stalled in the long living disk. 

Fig \ref{cold_gas_maps} shows the column density of the cold gas projected in a plane parallel to the jet axis at six different epochs, in a time window of $\approx700\ \mathrm{Myr}$ (from left to right), of the various runs $\kin$, $\kt$, $\ktnp$, $\tk$ and $\tknp$ (from top to bottom).
This figure has to be analysed jointly with the solid lines of Fig. \ref{cold_mass}, which show the amount of the cold gas mass formed (that is not accreted in the SMBH) in each run (right panel: runs with precession, left panel: runs without precession)  in the central volume of side $150\ \kpc$.
The run $\kin$ (first row of Fig. \ref{cold_gas_maps}) is one of the runs that form most cold gas, with a maximum of about $2.5\times10^{10}\ \msol$.
The  cold gas structure extends out to $100\ \kpc$ and $50\ \kpc$ along and perpendicularly to the jet axis, appearing as several wide filamentary structures composed of cold clumps raining towards the cluster center. 
This run forms a stable cold disk after $4\ \Gyr$, that continues accreting  until it has reached, $2\ \Gyr$ later, a total mass of $1.3\times10^{10}\ \msol$.
The latter  is in turn associated with a decreasing power of the jet: the cold circumnuclear gas tends to accrete onto the rotating disk, rather than feed the SMBH.

The run $\kt$ forms a cold disk already soon after the first episode of feedback ($\approx1\ \Gyr $), and it continues accreting gas during its whole evolution, but it also allows the prolonged accretion of gas onto the SMBH, which overall results into a gentle power history.
In this run, there are very rare episodes of cold clumps, and they do not out to a large radius. 
The disk which accretes continuously throughout the simulation reaches a total mass after $6\ \Gyr$ of $2\times10^{10}\ \msol$ and a size of $10-15\ \kpc$.
The run $\ktnp$ shows recurring episodes of clumps formations resulting sometimes in structures of size $100\ \kpc$ along the jet axis, but  smaller perpendicularly to the jet axis compared to run $\kin$. 
Overall, it does not form a huge amount of cold gas, having peaks around $6\times10^9\ \msol$.
The runs $\tk$ and $\tknp$ produce cold clumps intermittently, in the central region, during the whole evolution, mostly created by a continuous disruption of the disk when it forms.
In this case, the cold gas is more localized at the center in a region of size around $40\ \kpc$.

Fig.\ref{cold_mass} also gives some spatial information about the cold gas distribution: the solid lines measure the cold gas mass of the gas (not accreted on the SMBH) in a central volume of side $150\ \kpc$ while the dashed lines show the same quantity computed in a central volume of side $50\ \kpc$. 
We note that in runs $\kt$, $\tk$, and $\tknp$ the lines overlap, meaning that all the cold gas is within the central smaller region, while for the run $\kin$ and $\ktnp$, there are few epochs in which the $20-30\%$ of the cold gas is on larger scales.
For this reason, we can compare without ambiguity the kinematic of this phase with the hot phase data obtained in a smaller region.

It is clear now that some runs are more efficient than others in creating cold clumps and accreting them or in breaking the cold disk when it forms.
We can relate this trend with the AGN feedback modes and the presence of precession in a qualitative way.
Following the ideas of the cold gas formation driven by entrainment of the cold gas by the jets and a formation driven by turbulence at the edge of the jets, we could expect that the kinetic runs, promoting both these processes are the most efficient in creating cold gas.
In run $\kin$ the precession helps the turbulence created at the edge of the jets that in turn helps the clump formation. 
The runs $\tk$ and $\tknp$ having a little kinetic part, do not show a large amount of cold gas formation, but still, the precession helps the process since run $\tk$ forms more cold gas than $\tknp$.
The absence of precession in the runs with a thermal feedback component discourages the formation of cold gas: these feedback modes heat the ICM mostly in the same regions favouring long lasting hot bubbles at the same points, preventing the processes of turbulent mixing at their edges and consequent thermal instabilities and cold gas condensation. 
Indeed, the run $\ktnp$ with an important kinetic part produces more cold gas than $\tknp$, but still less than run $\tk$, which has a minor kinetic part but precessing. At variance with this is run $\kt$, in which the cold gas disk that rests untouched for all the simulation accretes all the cold gas that is formed in the run.

\subsubsection{Evolution of velocity dispersion}
The right panels of Figs. \ref{sigma_xyz_hotCold} show the velocity dispersion of the cold gas versus time, in the different runs computed in a box of size $150\ \kpc$, computed using Eq. \ref{eq_sigma_map}, counting only the cells with $T<5\times10^4\ \rm K$.
The velocity dispersion of this phase has a very oscillating behaviour with maximum values different for each run.
A correlation with the AGN activity is hard to find, although in the run $\kin$ it seems that $\sigma$ reaches large values before the AGN activity peaks.

The cold gas in the kinetic run $\kin$ has velocity dispersion values larger with respect to the other runs, as the hot phase does.  
In particular, it has maximum values around $300-500\ \kms$ for a short time.
This is due to the cold clumps that form at a certain height close to the vertical axis, and to their motion towards the center, gravitationally driven.
After $4\ \Gyr$, the cold gas disk is fully formed, and at this point, the velocity dispersion is very low.
In the $\kt$ run, the cold disk is formed after $1\ \Gyr$ and so the cold gas kinematic is dominated by that, showing low values of $\sigma$. 
The run $\ktnp$ shows a cold gas velocity dispersion with the major part of maximum values around $100\ \kms$ and some peaks around $200-300\ \kms$, showing no correlation at all with the AGN activity.
The runs $\tk$ and $\tknp$ have large values of velocity dispersion reaching $200-300\ \kms$, but the run $\tknp$ has some maximum peaks of $500\ \kms$, in the $y$ and $x$ directions. 

Looking at both panels of figure \ref{sigma_xyz_hotCold}, and comparing the hot and cold phase velocity dispersion, we conclude that the two phases are not significantly coupled.
Although the values range around the same
numbers sometimes there can be discrepancies of the $30-50\%$.

\section{Results: Velocity structure functions}\label{sec:vsf}
In this section, we analyze the properties of turbulent gas motions by computing the velocity structure functions (VSF) of the hot and cold phases in the various runs. 
Structure functions of the velocity field  are usually employed (together with the complementary view of power spectra) to study other turbulent astrophysical environments \citep[e.g.][]{2007ApJ...665..416K}.   
A few recent studies investigated turbulence in simulated galaxy clusters using VSF and reported statistics of velocity fluctuations compatible with the Kolmogorov model of turbulence\citep{Kol41}, i.e. $\mathrm{VSF} \propto l^{1/3}$ for the first order structure function, or slightly steeper \citep[e.g.][]{va09turbo,2020ApJ...889L...1L, Wang2021, Mohapatra22,  2022A&A...658A.149S}.  Observations of cold, ionized ICM return detailed information on the VSF, showing that its slope is always steeper than the Kolmogorov prediction, suggesting that the dynamics of cold gas reflects more the feedback driver than the true turbulent cascade \citep{li20,Hu22,Ganguly23}. It is therefore not clear if cold gas is a reliable tracer of hot gas turbulence \citep{Gaspari18}.

Here  we consider the first order VSF in the $i$-direction, defined as 
\begin{equation}
\mathrm{VSF}_i(l) =\langle |\mathrm{v}_i(\textbf{x}+{\textbf l}) - \mathrm{v}_i(\textbf{x})|\rangle.\label{vsf_true}
\end{equation}
The angular brackets $\langle\rangle$, which theoretically means the ensemble average, impossible to have in observations, here denote a spatial average.  
The computation is done in a box of side $150\ \kpc$ for both phases.
One of the objectives of the analysis is to determine the relation between the observed slope of the VSF, which is intrinsically emission-weighted projected along the LOS, and the real slope of the VSF in 3D, which is strongly linked to the physical properties of the turbulent evolution in the central cluster region. 

To infer that, we need the VSF computed using the emission-weighted projected velocities on the plane of the sky, obtained with Eqs. \ref{vproj_hot} and \ref{vproj_cold} for the hot and cold phase respectively, in Eq. \ref{vsf_true}.
In that way, we obtain an emission-weighted VSF along the LOS, $\vsfl$.
We compute $\vsfl$ along the three coordinate axes $x$, $y$, orthogonal to the axis of the jets, and $z$ parallel to that axis.  
Then we compare the $\vsfl$ with the intrinsic 3D VSF computed using Eq. \ref{vsf_true} substituting the $i$ component of the velocities considering for the hot phase only cells with a temperature above $10^6\ \rm K$ and for the cold gas cells with $T<5\times10^4\ \rm K$, called so on $\vsft$.

\begin{figure}
\begin{center}
\includegraphics[width=\linewidth,height=\textheight,keepaspectratio]{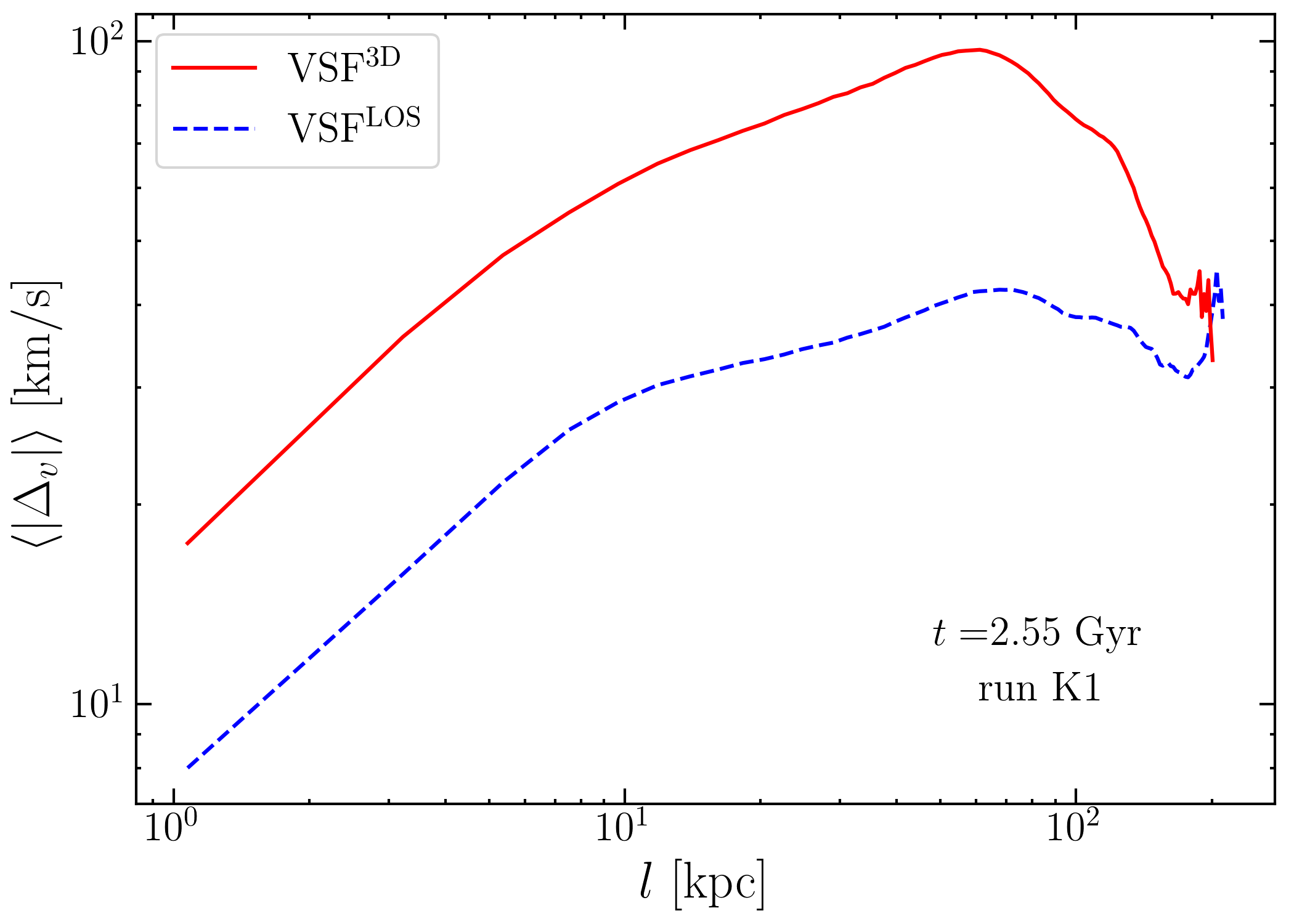}
\caption{Example of a $\vsft$ (solid-red) and a $\vsfl$ (dashed-blue) computed for a snapshot of the run $\kin$. }\label{vsf_example}

\end{center}
\end{figure}

\subsection{Hot phase VSF}
\begin{figure*}
\begin{center}
\includegraphics[width=\linewidth,height=\textheight,keepaspectratio]{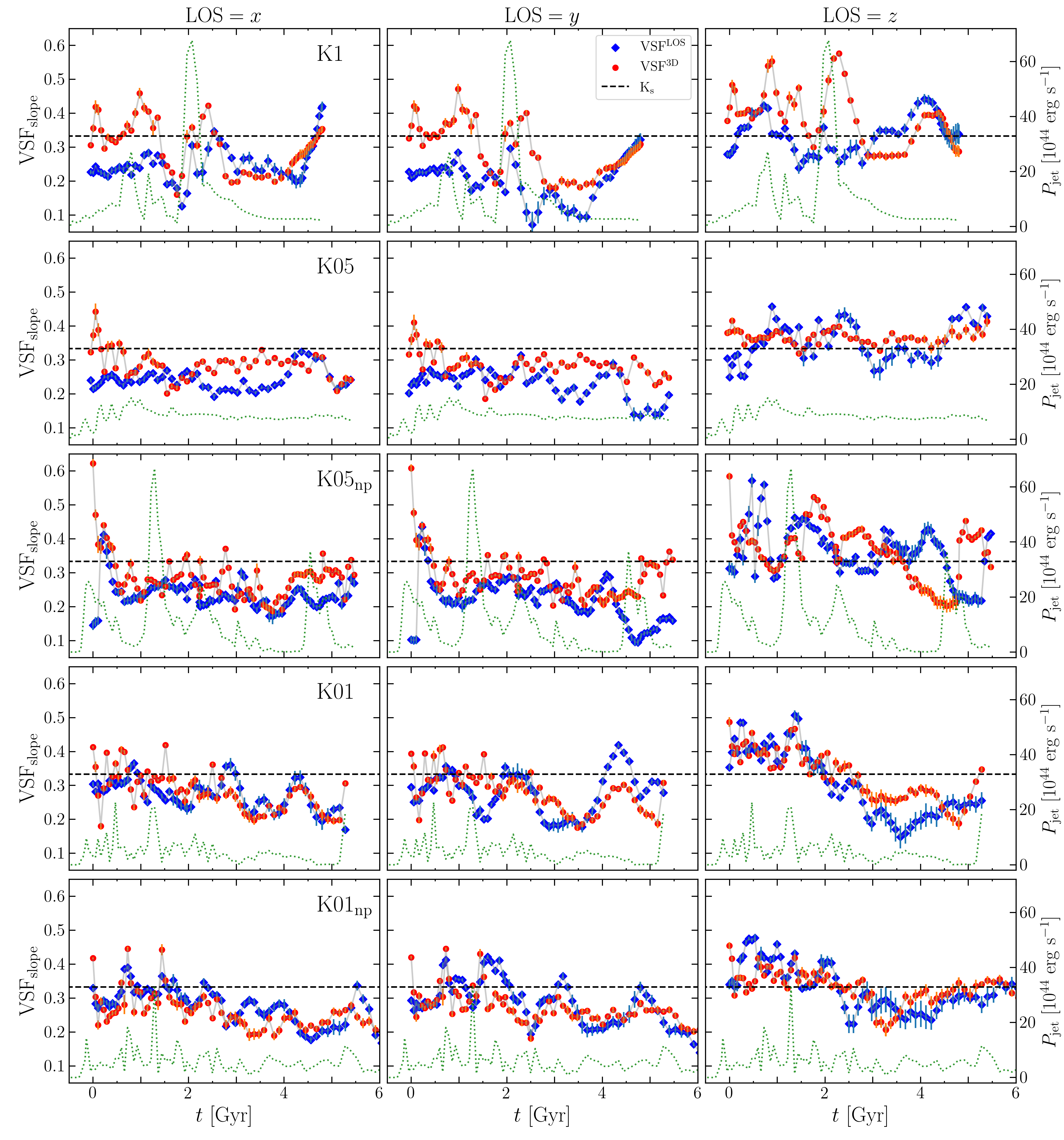}
\caption{Slopes of the hot gas 3D velocity structure function, $\vsft$ (red circles), and projected one, $\vsfl$ (blue diamonds), of the $x$, $y$ and $z$ components of velocity and corresponding projection direction (from left to right) fitted in the range $10-80\ \kpc$, of the run $\kin$, $\kt$, $\ktnp$, $\tk$ and $\tknp$ (from top to bottom). The green dotted lines show the AGN power, while the dashed black line is the Kolmogorov slope.} 
\label{vsf_slope}
	\end{center}	
\end{figure*}

As an example, we show in Figure \ref{vsf_example} the  VSF for model K1 computed at $t=2.55\ \Gyr$.
We fit the slope of the VSF in a scale range $l\approx[10-80]\ \kpc$.
The peak occurring at scale $\sim 70$ kpc is likely due to the finite box effect, rather than the physical scale of the turbulence driver (see discussion in Li et al., submitted).
At scales $\lesssim 10$ kpc the slope steepens (this is quite a general property of our models), with slope $\sim 2/3$. However, these scales are close to the simulation resolution (0.25 kpc) and it is unclear how much they reflect a real physical effect.

Figure \ref{vsf_slope} shows the $\vsft$ (red circles) and $\vsfl$ (blue diamonds) slopes of the $x$, $y$ and $z$ direction (columns from left to right), of the various runs $\kin$, $\kt$, $\ktnp$, $\tk$ and $\tknp$, (rows from top to bottom), computed on a scale range $l\approx[10-80\ \kpc]$.
The green dotted line shows the jet power of each run and the black dashed line is the Kolmogorov turbulence slope $1/3$.
We find no obvious general correlation or trend between the $\vsft$ and $\vsfl$ slopes.
For most models, and most of the time, the two slopes are very close. 
In the kinetic $\kin$ model there is a tendency for the $\vsft$ slope to be steeper than the $\vsfl$ one by $20-30$ \%.  
Some runs ($\kin$, $\tk$ and $\tknp$) show a tendency to have the peaks of the two VSFs in phase, but in several cases (first $3\ \Gyr$ run $\kin$) the peaks in the $\vsft$ slopes, i.e. the steepening of the slopes, disappear in the projection, while in other cases (the final part of $\kin$ along the $z$ axis and the $x$ and $y$ VSF of runs $\tk$ and $\tknp$), some steepening of the slopes are enhanced in the projections. 
A generally expected result is that, since the ICM dynamics is not steady throughout the simulation, the slopes in all models vary with time showing an oscillating behaviour.  
Looking carefully, we note that the $\vsft$ slope gets steeper after an AGN outburst.
We can explain these phenomena considering that the AGN bursts, especially the powerful ones, put in motion gas in larger regions creating bigger vortices and thus increasing the VSF amplitude at bigger scales.
During periods of low activities, the VSF slopes become flatter, and trace more closely the turbulence cascade.
If we consider that an eddy life $\tau$, is of the order of its turnover time $\tau\approx v/l$, being $v$ and $l$ its velocity and size, an eddy of $80\ \kpc$ moving at $200\ \kms$, lives less than $500\ \mathrm{Myr}$.
In general, the VSF slopes oscillate around the Kolmogorov slope, but most of the time are flatter than it. This contrasts with observations of VSF for the cold phase (see references above).

\subsection{Cold phase VSF}
\begin{figure*}
\begin{center}
\includegraphics[scale = 0.72]{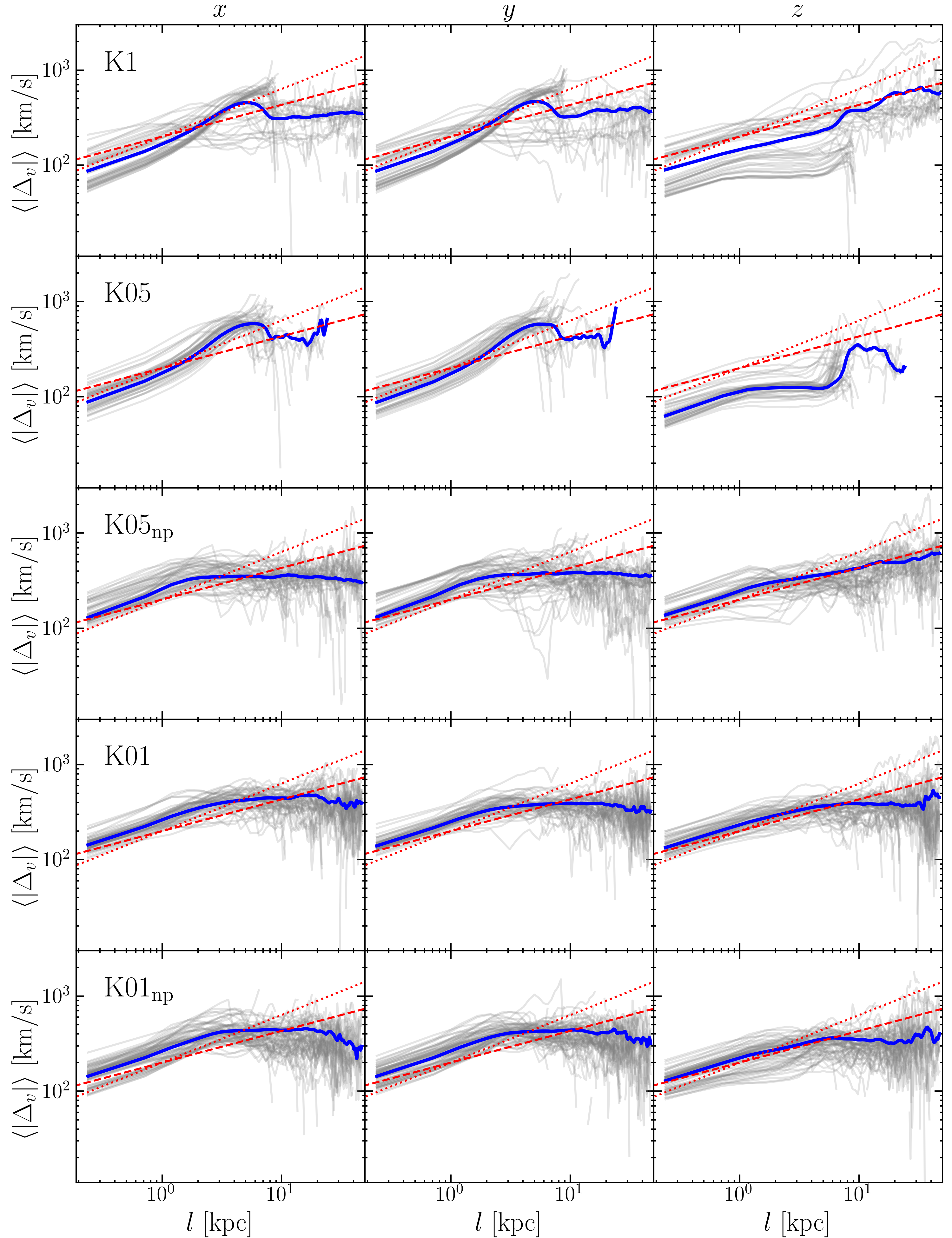}
\caption{Cold gas phase $\vsfl$ of the $x$, $y$ and $z$ components of velocity and corresponding projection direction (from left to the right), taken at all the snapshots (grey lines) of the run $\kin$, $\kt$, $\ktnp$, $\tk$ and $\tknp$ (from top to the bottom). 
The blue lines show the time average of all the snapshots. The dashed and dotted red lines indicate a slope $1/3$ and $1/2$ respectively.}
\label{vsf_cold}
	\end{center}	
\end{figure*}

For the cold phase, we compute the $\vsft$ and the $\vsfl$ based on the projected velocity obtained with Eq. \ref{vproj_cold}.
With respect to the hot phase, here we have fewer statistics due to the fact that the cold phase is not volume filling, but clumpy and located in small substructures.
For this reason is not possible to have for all the snapshots a VSF for which is possible to measure a slope.
Thus, we follow \citet{Wang2021} and average the VSF in time, in order to have enough data to infer some statistical properties.
Figure \ref{vsf_cold} shows the $\vsft$ average during all the time in which the cold gas is present, overplotted on each specific VSF.

Again, we find that the cold gas VSF is not steady, but that it reflects the intermittent nature of the AGN feedback. 
Both the slope and the amplitude vary in time. 
From the ensemble of grey curves, i.e. of the single $\vsft$s, is evident that after a certain spatial scale the VSFs become too noisy and measuring a slope is impossible.
However, we can note that when the VSFs are regular, within a spatial scale varying in time  and hardly exceed several $\kpc$, the slopes are steeper than Kolmogorov \citep[as in][]{li20,Wang2021}.

We can also find the moments in which the rotation, that occurs especially around the jet axis, is the dominant kinematic characteristic.
Since in the run $\kt$ the cold disc is present for most of the time, the resulting VSFs show this physical behaviour.
Along the $x$ and $y$ directions, the VSF are very steep with slope $\lesssim1$ and on the $z$ direction the VSF is flat.
This happens in the final $\Gyr$ of run $\kin$ and can be spotted
for short periods in the other runs.

Concerning the largest scales the time average of the VSF seems to suggest a flattening of the slope.
The explanation for the single VSF can be found considering that the cold gas is clumpy and spatially concentrated.
Especially the external parts are composed of a small number of clumps, thus the big scales are poorly sampled, given the small number of pairs, resulting in a noisy VSF.
Moreover, as we found for the hot gas, the effect of the finite box in which the VSF is computed can give a flattening of the VSF at scales close to the half of the box.
In the case of cold gas, the effect of the finite box is natural since the gas is naturally confined in small regions.

On the other hand considering the noise of all the VSFs combined and the fact that they come for different kinematic behaviour of the gas, absolutely not steady, we do not trust the sum of all the pairs as reflecting the intrinsic VSF characteristics.
Thus the average procedure, even though gives a regular VSF, is not trustworthy as a tracer of the turbulent motion in the cold phase.

\section{Discussion and conclusions} \label{sec:conclusions}
In this work we have produced new hydrodynamic simulations of AGN feedback in a Perseus-like cluster of galaxies,to study the impact of different prescriptions for feedback on the kinematics of the ICM. We investigated models where the BH injects energy with various combinations of kinetic and thermal components. Moreover, we explored the effect of the AGN jet precession on the ICM properties.
 
A first important finding is that the kinematics of the ICM is strongly correlated with the AGN activity, which is in turn 
regulated by the  feedback mode, and its efficiency in creating cold gas.
The accretion of cold gas controls the feedback power, setting the history of the AGN power, which can result in a bursty, intermittent or gentle regime. In this context, the presence of precession changes the behaviour of the AGN, especially for the models in which the feedback energy is mainly deposited as kinetic energy.

The time evolution of the velocity dispersion 
of the hot and cold gas phases, $\sigma_{\rm hot}$ and $\sigma_{\rm cold}$, (Fig. \ref{sigma_xyz_hotCold}) shows that for the kinetic mode runs it exhibits strong variations, roughly in pace with the AGN activity.
However, the peaks of $\sigma_{\rm cold}$ often precede the bursts in the AGN activity, by $\approx 10^8$ yr, which corresponds to a few dynamical times at the location of the cold gas. 
Evidently, large values for $\sigma_{\rm cold}$ are the signature of the infall of cold clouds toward the central BH, where they are eventually accreted.
In contrast to $\sigma_{\rm cold}$, $\sigma_{\rm hot}$ is tightly correlated with the peaks of AGN power, and this happens for all the models presented here, regardless of the kinetic/thermal energy fraction.
This overall agrees with the general picture of cold gas feeding the SMBH and causing the hot gas stirring \citep{gaspari12}.

However, the velocity dispersion analysis shows that, in general, cold and hot gas velocity dispersion do not perfectly trace each other, but a discrepancy up to a factor of two is common. 
Besides the peak separation in time observed in model K1, $\sigma_{\rm cold}$ displays larger time variations than $\sigma_{\rm hot}$, an effect that we ascribe to the dominant effect of gravity on the cold clumps. Finally, the time averaged value for $\sigma_{\rm cold}$ is generally lower than $\sigma_{\rm hot}$.
The latter result is in contrast with the finding of \citet{valentiniBrigh15}, who found $\sigma_{\rm cold} > \sigma_{\rm hot}$ (see their Fig. 11), and \citet{Gaspari18}, who found instead $\sigma_{\rm cold}\sim\sigma_{\rm hot}$, albeit with a dispersion of $30-40$ \% (their Fig. 1). 
Clearly, further investigation must be carried out to clarify the relation between the motion of the various thermal phases of the ICM.

A robust result of our simulations is that the central hot gas velocity dispersion (Fig. \ref{sigma_xyz_hotCold}) agrees with the one observed at the center of Perseus. 
For more than half of the evolutionary time, we found that $\sigma_{\rm hot}$ lies in the observed range (colored band in Fig. \ref{sigma_xyz_hotCold}, \citealt{hitomi16}), and this is true for every model presented here.
Evidently, the self-regulated AGN activity alone is able to stir the ICM, irrespective of the detail of the feedback nature and numerical implementation. 

Further information is gained through the analysis of $\sigma_{\rm hot}$ maps (see Fig. \ref{fig:sigma_pro}). 
The radial profiles along several directions show that different feedback models lead to different geometrical distributions of the velocity dispersion in the cluster core. 
The magnitude of the velocity dispersion radial drop depends on both the feedback mode and the direction with respect to the LOS. 
In particular, our calculations show that $\sigma_{\rm hot}$ is about constant along the outflow direction if the jet is seen almost edge-on. 
This result can  be used to verify the presence of large scale, non-relativistic hot gas outflows as a major AGN heating mechanism. 
Forthcoming data by XRISM telescope will allow spatial analysis of this kind in the brightest cluster cores and will offer observational tests to discriminate among feedback modes, as well as to identify the presence of jet precession. 

A more quantitative analysis of the dynamics and turbulence of the multiphase ICM uses the (first order) velocity structure functions.
Our runs show that for the hot gas, the slopes of the projected and the intrinsic VSF are similar, but most of the time the $\vsft$ is slightly steeper than the projected one. 
In a few cases, however, the opposite happens (Fig. \ref{vsf_slope}).

This is in contrast with what is expected theoretically, as derived by \cite{xu20}, who predicted always a steeper slope, due to projection.
On the other hand \citet{Mohapatra22}, with a different implementation of turbulence recovered both the effects. 
We think that in our case, the lack of stationary turbulence, the extreme variability of the driving scale as well as of the driving power, combined with the episodic strong outflows and anisotropy of the turbulent cascade, makes our turbulence system very far from homogeneous and isotropic. This implies that the VSFs are often not single-slope and it is impossible to derive a clear relation between 3D and projected VSF slopes.

This is more evident in the runs with dominant kinetic feedback that injects turbulence into the ICM  anisotropically; the $\vsft$ are always steeper than Kolmogorov during the strong outbursts.
The steepening of the VSF can be due to AGN-driven coherent outflows and an excess of turbulent eddies at larger scales, combined with a scarcity of small-scale turbulent eddies. 
Especially for the kinetic run, $\kin$, the steepening in the intrinsic VSF during the strong outburst can be due to the fact that the AGN, in those epochs, is injecting turbulence on large scales, breaking the ideal Kolmogorov cascade model (if present at all). 
Thus, after the AGN burst, there is a relative excess of large-scale eddies, which have not had time yet to be processed by the Kolmogorov cascade. 
Therefore, during these moments the dominant projection effect is the cancellation of the large-scale fluctuations due to the apparent smaller distances, leading to a flattening of the VSF.

When the AGN is in a low activity state, the injected turbulence can evolve through the full cascade.  When this happens, the discrepancy disappears and the projected VSFs become steeper than the 3D ones as theoretically predicted.
In the end, we found that there is no obvious quantitative connection between the intrinsic and the projected  VSF slopes.  In principle, although outside of the goal of this paper, further projects can be made by means of controlled experiments, in which the AGN feedback power is kept exactly constant among models so that all VSFs are computed for the same overall amount of energy input. 

Moreover, we found that most of the time the VSF slope is close to the theoretical Kolmogorov one ($1/3$), with a tendency to be slightly steeper than Kolmogorov when the VSF is computed in the direction parallel to the AGN jet axis and slightly flatter when computed in the direction perpendicular to the jet axis.

The study of the cold gas VSF is of particular importance, given the wealth of data available via integral field unit spectrographs, both in the optical and submillimeter bands.
This is a difficult task for simulations because the internal structure of the cold gas is likely not fully resolved. Our results (Fig. \ref{vsf_cold}) suggest that performing a  time average in all our runs does not yield a slope that corresponds to the true VSF measure at a specific time, which can have a broad range of stochastic behaviours.
The single cold gas VSF slopes have a quasi-Kolmogorov slope ($\sim 1/3$) in the scale range $1-5$ kpc and flattening for larger separations, a likely consequence of the finite size of the cold gas system (typical size $\approx 10$ kpc).
Observed VSF for ionized and molecular gas in BCGs show a steeper than Kolmogorov slope \citep{li20}, but observations also suffer from projection and smoothing effects which are not easily quantifiable (see \citealt{Chen23} and the discussion in Li et al., submitted).

To conclude, we summarise our main results as follows: 
\begin{itemize}[label={$\bullet$}]
    \item In self-regulated AGN cycles, different feedback modes and the presence of precession lead to a different efficiency in cold gas formation, which in turn drastically changes the power history of the AGN. As a consequence, also the gas velocity dispersion and the VSFs of the hot and the cold gas phases are different in all cases, suggesting potential pathways to discriminate between different AGN feedback modes. 
    \item The AGN feedback with different modalities and characteristics is able to produce the hot gas velocity dispersion measured at the center of the Perseus cluster.  The powerful test that may constrain which feedback mode is at play is the measurement of velocity dispersion profiles across the central region.
    \item Our explored variations of  implementations of AGN  feedback  show that the cold and hot phases are not tightly correlated, with typical discrepancies of the $20-30\%$ in values of the velocity dispersion of the two phases.  The observed time lag between the turbulent statistics of both phases suggests that the cold gas kinematics is more associated with an inflow of the gas while the hot gas kinematics is associated with the outflow from the AGN.
    \item The VSF slopes of the hot gas are sensitive to the AGN activity as they get steeper during strong AGN outbursts, especially for the ones computed along the direction parallel to the jet axis.
    \item The VSF slopes, excluding the epoch in which the AGN outbursts make them steep, are close to the Kolmogorov values showing a tendency to be slightly flatter when computed perpendicularly to the jet axis and slightly steeper when computed parallel to the jet axis.
    \item There is no trivial trend between the intrinsic VSF and the projected one, however, the oscillations in the slopes experienced are in phase.
    \item The cold gas VSFs have a slope steeper than Kolmogorov value, $\sim1/2$, at small separations $l < 5$ kpc and get flatter at larger scales, likely because the finite size of the emitting region.
    \item The time average of the VSF is not a good procedure due to the non steady state of the motion of the phases.
    
\end{itemize}
In the future, we plan to test other combinations of feedback modalities and precession properties improving the AGN feedback implementation and adding missing physics as colder gas phases.
At the same time, multi-wavelength observations are required for a deeper understanding of these complex processes.

\section*{Acknowledgements}
F.V. has been supported by Fondazione Cariplo and Fondazione CDP, through grant n° Rif: 2022-2088 CUP J33C22004310003 for "BREAKTHRU" project.

\bibliographystyle{aa}
\bibliography{biblio_Ste}

\end{document}